\documentclass[floatfix,preprint,superscriptaddress,amsmath,amssymb,aps,showkeys,titlepage]{revtex4-2}

\usepackage[utf8]{inputenc}
\usepackage[english]{babel}

\usepackage{graphicx}
\usepackage{subcaption}
\usepackage{dcolumn}
\usepackage{bm}

\usepackage{caption}

\usepackage{ragged2e}

\DeclareCaptionJustification{justified}{\justifying}

\usepackage[colorlinks = true,
            linkcolor = blue,
            urlcolor  = blue,
            citecolor = red,
            anchorcolor = blue]{hyperref}

\begin{document}

\title{Reversible Modulation of Thermal Conductivity in GaN through Strain-Field Screening around Dislocations}

\author{Shantal Adajian}
\affiliation{Department of Mechanical Engineering, University of California, Santa Barbara, CA 93106, USA}

\author{Fanghao Zhang}
\affiliation{Department of Mechanical Engineering, University of California, Santa Barbara, CA 93106, USA}

\author{Andrew Christison}
\affiliation{Department of Mechanical Engineering, University of California, Santa Barbara, CA 93106, USA}

\author{Zeyu Xiang}
\affiliation{Department of Mechanical Engineering, University of California, Santa Barbara, CA 93106, USA}

\author{Tanay Tak}
\affiliation{Materials Department, University of California, Santa Barbara, CA 93106, USA}

\author{Nicholas Fuchs-Lynch}
\affiliation{Materials Department, University of California, Santa Barbara, CA 93106, USA}

\author{Szilvia Kalácska}
\affiliation{Mines Saint-Étienne, Univ Lyon, CNRS, UMR 5307 LGF, Centre SMS, 158 cours Fauriel, 42023, Saint-Étienne, France}

\author{Nicolò M. della Ventura}
\affiliation{Materials Department, University of California, Santa Barbara, CA 93106, USA}

\author{Miguel Zepeda-Rosales}
\affiliation{Materials Research Laboratory, University of California, Santa Barbara, CA 93106, USA}

\author{Samantha Daly}
\affiliation{Department of Mechanical Engineering, University of California, Santa Barbara, CA 93106, USA}

\author{Irene J. Beyerlein}
\affiliation{Materials Department, University of California, Santa Barbara, CA 93106, USA}
\affiliation{Department of Mechanical Engineering, University of California, Santa Barbara, CA 93106, USA}

\author{Bolin Liao}
\email{bliao@ucsb.edu}
\affiliation{Department of Mechanical Engineering, University of California, Santa Barbara, CA 93106, USA}

\date{\today}

\begin{abstract}
Crystalline defects are generally regarded as static phonon scatterers that irreversibly suppress thermal transport. Here we show that elastic strain can reversibly modify dislocation-associated strain fields and strongly alter heat conduction. Using \textit{in situ} strain-dependent time-domain thermoreflectance measurements, we observe a reversible enhancement of thermal conductivity in GaN by 23\% under only 0.21\% uniaxial strain. High-resolution X-ray diffraction reveals progressive narrowing of the symmetric (0002) reflection, indicating a reduction in the distribution of lattice rotations and heterogeneous strain. High-resolution electron backscatter diffraction directly shows that the spatial autocorrelations of multiple strain components decay over progressively shorter distances with applied strain, providing real-space evidence for enhanced screening of long-range strain fields. Raman spectroscopy further shows a non-monotonic evolution of the $E_{2}^{\mathrm{high}}$ phonon linewidth near the onset of the thermal-conductivity increase. Together, these results support a picture in which elastic strain reversibly reconfigures pinned dislocation lines and shortens the spatial range of their heterogeneous strain fields, thereby reducing phonon scattering. Our work establishes defect-associated strain correlations as a tunable degree of freedom for controlling thermal transport in crystalline solids.
\end{abstract}

\maketitle

\section{Introduction}

Thermal transport in crystalline insulators and semiconductors is governed by the propagation and scattering of phonons. In these materials, lattice thermal conductivity is highly sensitive to structural disorder, including point defects, interfaces, grain boundaries, and dislocations that effectively scatter phonons and disrupt thermal transport~\cite{hanus2021thermal}. Among these, dislocations are particularly important because they combine a localized core with long-range elastic strain fields that perturb phonons over distances far exceeding the atomic scale~\cite{klemens1955scattering,wang/2017/ab}. As a result, even moderate dislocation densities can substantially suppress thermal conductivity, especially in wide-bandgap semiconductors and heteroepitaxial thin films where threading dislocations are commonly introduced during growth~\cite{Kaganer2005GaNDislocations}.

Gallium nitride (GaN) provides an important model system for studying defect-limited heat transport. Owing to its wide bandgap, high breakdown field, and favorable electron transport properties, GaN underpins high-power electronics, radio-frequency devices, and optoelectronics. In many of these technologies, heat dissipation is a central performance bottleneck~\cite{cho/2015/near-junction,malakoutian/2021/record-low,meneghini/2021/gan-based,kuball/2016/review}, making the thermal conductivity of GaN technologically critical~\cite{choi2021perspective}. While bulk single-crystal GaN can exhibit high thermal conductivity, practical GaN materials often contain threading dislocation densities in the range of $10^{7}$--$10^{10}~\mathrm{cm^{-2}}$ even in device-quality films~\cite{shih/2015/ultralow,speck/1999/role}, depending on substrate choice and growth method. These dislocations are widely believed to degrade heat conduction through phonon scattering arising from both their cores and associated strain fields~\cite{MionAPL2006,park/2019/impact,wang2019phonon,li/2020/gan}.

Most existing strategies for mitigating defect-limited thermal transport rely on materials synthesis: reducing defect densities, improving crystal quality, changing substrates, or introducing post-growth annealing treatments~\cite{chen2021effects,niyikiza2025thermal}. Although effective, such approaches are static in nature. Once the crystal is grown, the dislocation population is typically regarded as fixed, and the associated thermal resistance is treated as an intrinsic materials limitation. This viewpoint overlooks the fact that dislocations are not merely isolated defects, but interacting elastic objects capable of collective behavior~\cite{zaiser2001statistical,el2000statistical}. Their long-range stress fields can attract, repel, align, screen, and reorganize under external stimuli~\cite{byun2003stress,zaiser2001statistical,li2023harnessing}, suggesting that the phonon scattering strength of a dislocation ensemble may be tunable even when the total defect density remains unchanged.

Mechanical strain offers a promising route to access this physics. It is often unavoidable during growth due to lattice-constant and thermal-expansion mismatches~\cite{kisielowski1996strain}, and it is also deliberately introduced to enhance electrical charge mobility~\cite{fischetti2002enhanced}. Despite its broad importance, strain-dependent thermal conductivity has received comparatively limited experimental attention. For example, our prior work showed that residual biaxial thermal stress in epitaxial GaAs films on Si can reduce the thermal conductivity of GaAs, which we attributed to enhanced phonon scattering associated with strain-induced symmetry breaking~\cite{vega2019reduced}. Other measurements on single crystals typically report only a modest decrease in the thermal conductivity under tensile strain~\cite{chen2023plane,murphy2014strain}, consistent with classical models and first-principles predictions~\cite{parrish2014origins}.A qualitatively different possibility emerges in defect-rich crystals: external strain may alter the statistical organization of dislocations themselves~\cite{zaiser2001statistical}. Because phonons are sensitive to the spatial distribution and correlation of strain fields, even subtle defect reorganization could strongly modify heat conduction. Despite this possibility, direct experimental evidence for strain-tunable phonon scattering mediated by collective dislocation behavior remains scarce.

Here we use GaN as a model platform to demonstrate reversible control of thermal transport through strain-induced changes in dislocation-associated strain fields. Using a suspended microdevice that enables \textit{in situ} uniaxial loading during time-domain thermoreflectance (TDTR) measurements, we observe a large and reversible enhancement of thermal conductivity by 23\% under only $0.21\%$ applied tensile strain. HRXRD reveals progressive narrowing of the (0002) reflection with applied strain. More directly, high-angular resolution electron backscatter diffraction (HR-EBSD) maps show that the spatial autocorrelations of the heterogeneous strain field contract strongly under loading, consistent with enhanced screening of elastic long-range dislocation strain fields~\cite{kaganer2025dislocation}. Raman spectroscopy provides a complementary local signature through the non-monotonic evolution of the in-plane $E_{2}^{\mathrm{high}}$ phonon linewidth. These observations support a picture in which external strain reversibly changes the geometry of pinned threading dislocations and the spatial range of their elastic fields without requiring the creation or annihilation of dislocations. Our results establish defect-associated strain correlations as a previously underexplored degree of freedom for tuning heat transport.

\section{Results and Discussion}

\subsection{Strain-dependent thermal conductivity in GaN}

\begin{figure*}[t]
  \centering
  \includegraphics[width=\textwidth,height=0.65\textheight,keepaspectratio]{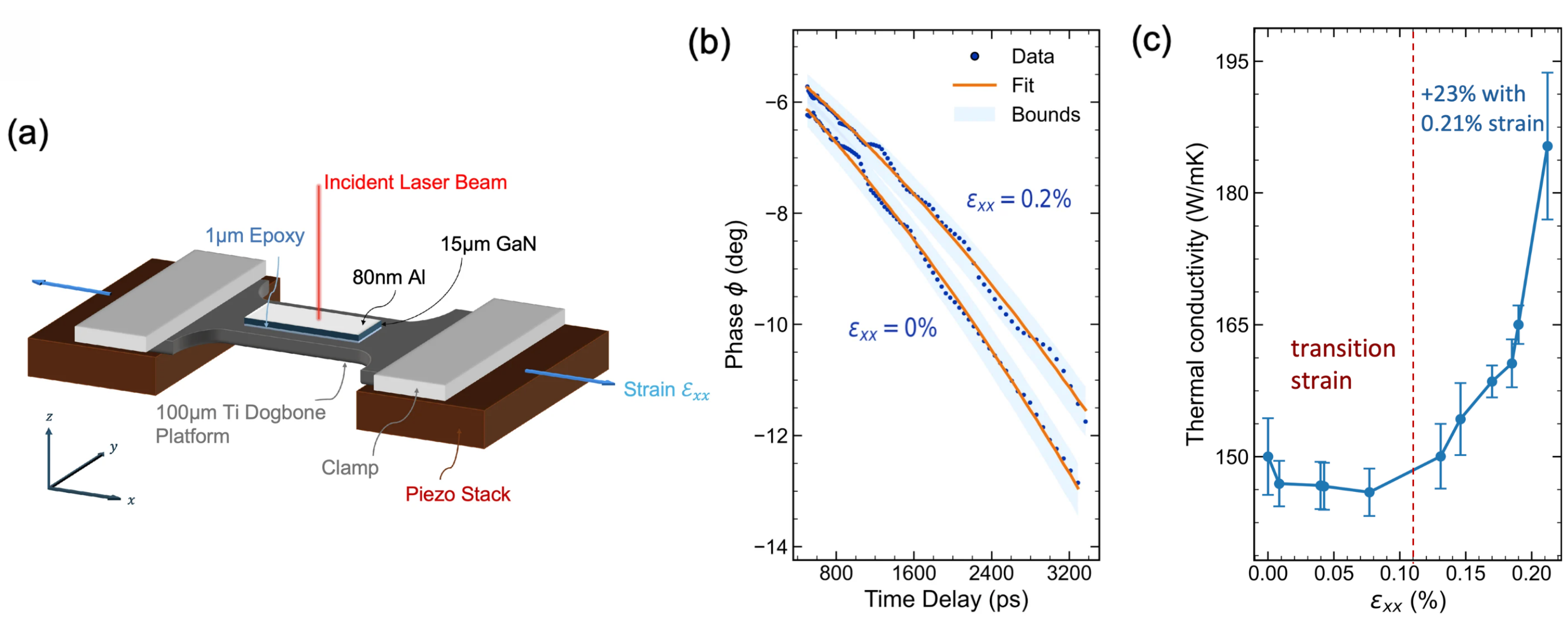}
  \caption{\textbf{In situ strain-dependent thermal conductivity measurements in c-plane GaN.} (a) Schematic of the experimental setup integrating an in situ uniaxial strain platform with TDTR. Tensile strain is applied using piezoelectric stacks that load a titanium dog-bone platform via mechanical clamps, while optical access is maintained for TDTR measurements. (b) TDTR phase signal versus pump-probe delay time at zero and 0.2\% tensile strain. (c) Measured cross-plane thermal conductivity as a function of applied in-plane tensile strain. Error bars represent the standard deviation from repeated measurements.}
  \label{fig:Figure_1}
\end{figure*}

To probe whether elastic strain can dynamically modify heat transport in defect-rich GaN, we integrated a suspended uniaxial strain platform with time-domain thermoreflectance (TDTR)~\cite{jiang2018tutorial}, as illustrated in Fig.~\ref{fig:Figure_1}(a). A mechanically thinned c-plane GaN single crystal ($15~\mu$m thick) was bonded onto a titanium substrate that transfers controllable in-plane tensile strain~\cite{park/2020/rigid} (see Supplementary Note 1) while preserving optical access for TDTR. Representative phase signals at zero and finite strain are shown in Fig.~\ref{fig:Figure_1}(b). Clear strain-dependent shifts are observed, while the data remain well described by the thermal model over the full delay-time range. Detailed TDTR data analysis is provided in Supplementary Note 2.

The extracted cross-plane thermal conductivity along the $c$ axis as a function of applied tensile strain along one direction within the $ab$ plane in GaN is summarized in Fig.~\ref{fig:Figure_1}(c). In the unstrained state, the measured conductivity is approximately $150~\mathrm{W\,m^{-1}\,K^{-1}}$, lower than that of pristine bulk GaN, likely due to the combined influence of pre-existing crystalline defects and thinning-induced damage or residual strain~\cite{li/2020/gan}. Because all measurements were performed on the same specimen and at the same location, the central quantity of interest is the reversible evolution with strain. Upon increasing strain, the thermal conductivity initially changes weakly below $\sim0.1\%$ strain, followed by a pronounced rise above this threshold. At a maximum applied strain of $0.21\%$, the conductivity increases by $23\%$. Repeated loading-unloading cycles reproduced the same TDTR response and extracted conductivity within experimental uncertainty (see Supplementary Note 3), indicating that the effect is elastic and reversible rather than damage-driven.

The magnitude and nonlinear onset of the conductivity enhancement are notable. For sub-percent elastic strain, intrinsic changes in phonon dispersion, group velocity, or phonon-phonon scattering are generally expected to produce only modest and smooth variations in thermal conductivity. First-principles calculations based on density functional perturbation theory and the phonon Boltzmann transport equation likewise predict a substantially smaller response (see Supplementary Note 4). The observed threshold-like enhancement therefore points to an additional strain-sensitive extrinsic scattering mechanism. Because TDTR probes several micrometers into the sample under the present conditions (10 MHz modulation frequency)~\cite{jiang2018tutorial}, the measured response primarily reflects bulk phonon transport rather than surface effects. We hypothesize that the anomalously large increase in thermal conductivity can be attributed to a strain-induced reduction in phonon scattering by extended defects, most notably threading dislocations. As shown below, HRXRD, HR-EBSD, and Raman measurements reveal complementary changes in lattice disorder and strain correlations, supporting a picture of strain-induced screening of dislocation-associated strain fields. As a control, we also measured m-plane GaN under uniaxial loading applied along both the $c$ and $a$ axes (see Supplementary Note 5). In contrast to the c-plane sample, neither configuration exhibited a clear strain-dependent change in cross-plane thermal conductivity. This orientation dependence further indicates that the large conductivity modulation is not an intrinsic elastic response of GaN, but instead depends on the coupling between applied strain, heat-flow direction, and the specific defect structures in the c-plane sample.
\subsection{Strain-dependent screening of heterogeneous strain fields in GaN}

\begin{figure*}[t]
  \centering
  \includegraphics[width=0.9\textwidth]{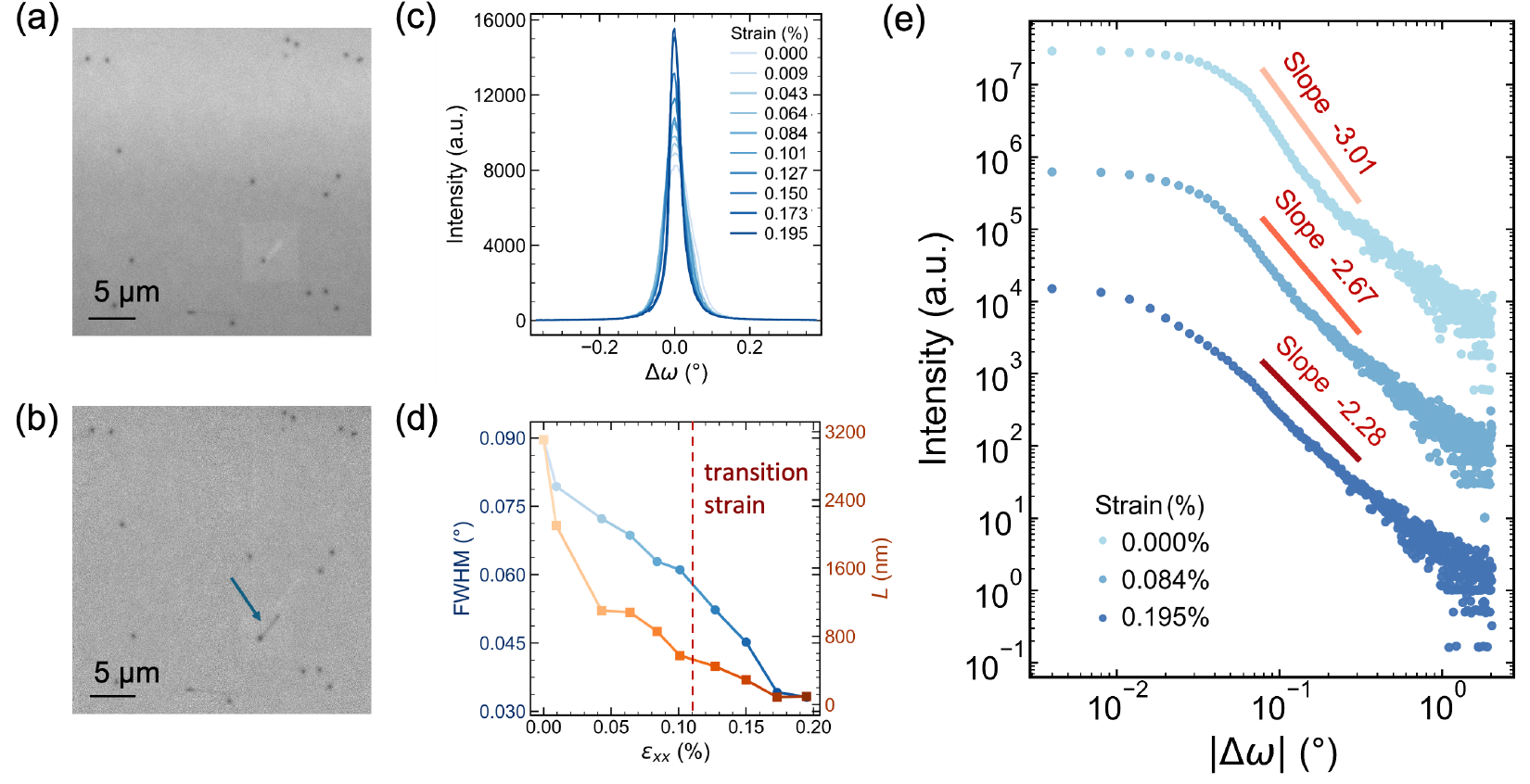}
  \caption{
  \textbf{Cathodoluminescence (CL) and high-resolution X-ray diffraction characterization of strained GaN.}
  (a) Plan-view CL intensity map acquired prior to strain cycling.
  (b) Plan-view CL intensity map acquired after strain cycling. The arrow highlights the only representative defect feature whose morphology evolved
  from a localized point-like contrast prior to straining into a short line-like feature after
  strain cycling.
  (c) In-situ GaN (0002) $\omega$-rocking curves measured at increasing applied tensile strain.
  The curves are aligned by their peak positions to account for small rigid-body sample
  motions during straining and to emphasize strain-dependent changes in rocking-curve
  shape. (d) The full-width-half-maximum (FWHM) of the rocking curve peak as a function of applied tensile strain, showing
  a systematic decrease in FWHM with increasing strain.
  (e) Tail-slope analysis of representative rocking curves. The inset summarizes the effective correlation range ($L$) obtained from full-profile fits and the measured tail slope as a function of applied strain.
  }
  \label{fig:Figure_2}
\end{figure*}

To investigate the microstructural origin of the thermal conductivity enhancement, we performed complementary cathodoluminescence (CL) imaging and \textit{in situ} high-resolution X-ray diffraction (HRXRD) measurements under comparable applied strain states. Figures~\ref{fig:Figure_2}(a) and (b) show plan-view CL maps acquired before and after repeated strain cycling, respectively. Within the spatial resolution of the CL measurements, no substantial change in the density, position, or overall contrast of surface-reaching threading dislocation outcrops is observed after cycling. This indicates that the applied loading does not generate new surface defects or induce large-scale irreversible plastic deformation, consistent with the reversible thermal-conductivity modulation described above. It is important to note, however, that CL primarily probes dislocation outcrops at or near the sample surface, whereas the morphology of threading dislocations within the bulk can be substantially more complex, including inclined segments, bends, and kinks~\cite{lu2007morphology}. Because the TDTR thermal penetration depth extends several micrometers into the sample, bulk dislocation configurations are expected to play a dominant role in determining the measured thermal transport response.

To probe the dislocation ensemble throughout the sample volume, we next performed \textit{in situ} HRXRD rocking-curve measurements of the symmetric GaN (0002) reflection, which is sensitive to screw dislocations and mixed dislocations with a screw component~\cite{heying/1996/role}. As shown in Fig.~\ref{fig:Figure_2}(c), increasing tensile strain produces a clear narrowing of the rocking curve together with an increase in peak intensity. The extracted full width at half maximum (FWHM), summarized in Fig.~\ref{fig:Figure_2}(d), decreases progressively with strain. Since lattice rotations and inhomogeneous strain associated with threading dislocations are well-known sources of diffraction broadening, the reduction in FWHM indicates a decrease in average lattice disorder and distortion~\cite{AYERS199471}. The penetration depth of the X-ray probe under the present geometry is on the order of micrometers, making HRXRD sensitive to the bulk rather than only the near-surface region.

To quantify the profile evolution beyond the FWHM, we fitted the full (0002) rocking curves using the restricted-random-dislocation model developed by Kaganer \textit{et al.}~\cite{Kaganer2005GaNDislocations} (Supplementary Note 6). In this framework, the diffraction profile is described by an effective dislocation density $\rho_{\mathrm{eff}}$ and a correlation range $L$, where $L$ represents the length scale over which dislocation strain fields remain unscreened. The fits yield an effective screw-character dislocation density of order $10^{9}~\mathrm{cm^{-2}}$ throughout the strain series, consistent with a density capable of influencing phonon transport in GaN~\cite{wang2019phonon,li/2020/gan}. The fitted $L$ decreases from approximately $3.1~\mu\mathrm{m}$ in the unstrained state to about $90~\mathrm{nm}$ at $0.195\%$ strain [inset of Fig.~\ref{fig:Figure_2}(e)]. This trend is consistent with a contraction of the spatial range of dislocation-related lattice distortions. We treat these fitted values as effective comparative metrics rather than absolute screening lengths because the measured far-tail exponent evolves away from the $q^{-3}$ asymptote of the model. The unchanged density of CL dislocation outcrops and the approximately constant effective screw-character density indicate that the XRD evolution does not require creation or annihilation of threading dislocations which are irreversible processes.

The diffuse tails evolve from approximately $I(q)\propto q^{-3}$ toward a shallower dependence near $q^{-2}$ [Fig.~\ref{fig:Figure_2}(e)]. This change demonstrates that the origin or relative weight of the broadening mechanisms changes with loading, but it is not by itself a unique signature of dislocation screening. In the Kaganer framework, the far-tail $q^{-3}$ asymptote is insensitive to the dislocation correlation range, and $q^{-2}$ behavior in symmetric reflections can also arise from finite-thickness or coherent-domain broadening~\cite{Kaganer2005GaNDislocations}. We therefore base the screening interpretation primarily on the direct real-space HR-EBSD correlations described below, while using the XRD narrowing and full-profile evolution as complementary volume-averaged evidence for reduced heterogeneous lattice distortion.

\begin{figure*}[t]
  \centering
  \includegraphics[
    height=0.52\textheight,
    width=0.9\textwidth,
    keepaspectratio,
    clip
  ]{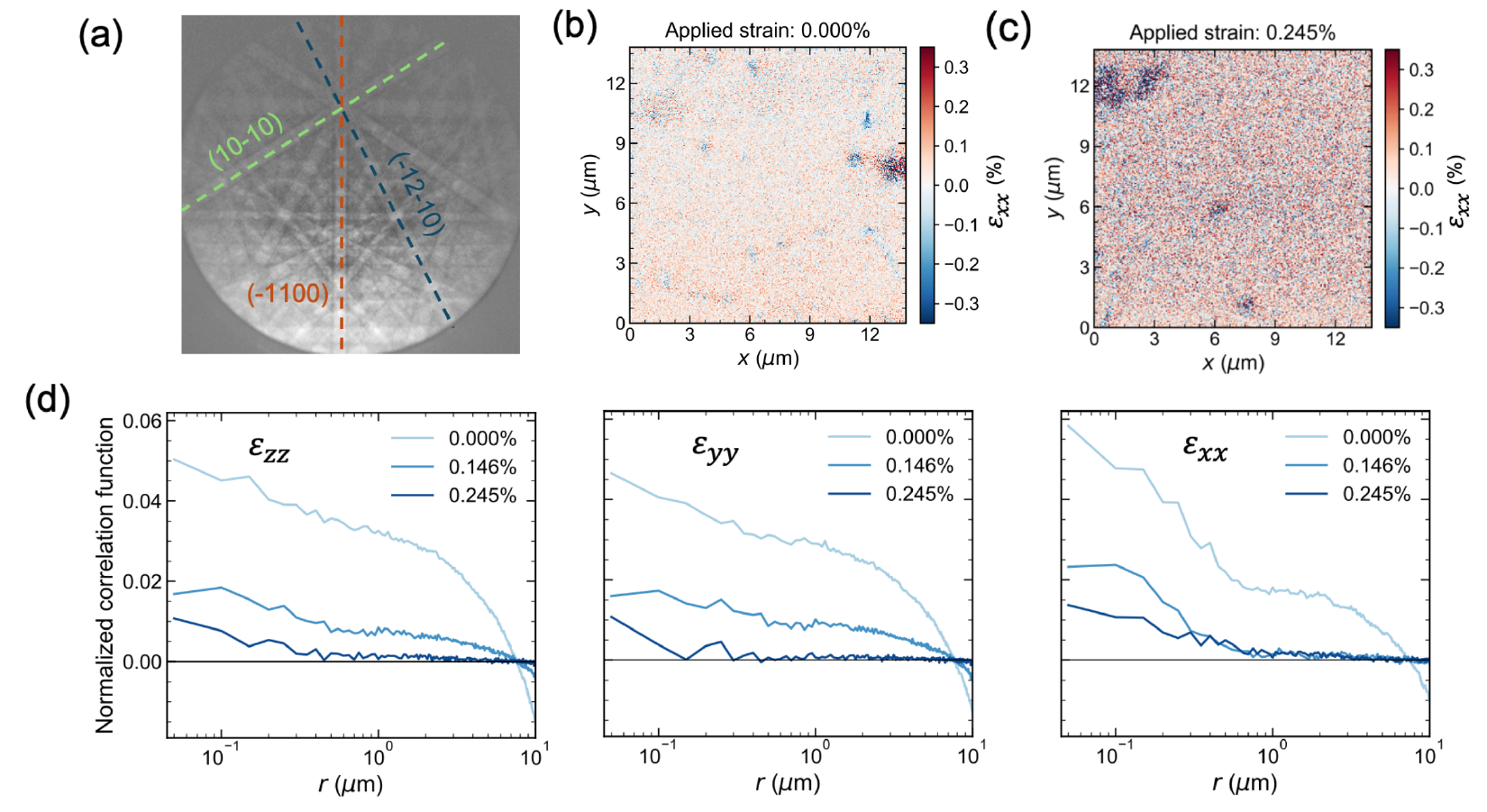}
  \caption{\textbf{Real-space strain correlations measured by high-resolution electron backscatter diffraction.}
  (a) Representative Kikuchi pattern used for cross-correlation analysis.
  (b,c) Maps of $\varepsilon_{xx}$ at applied macroscopic strains of 0 and $0.245\%$, respectively.
  (d) Radially averaged autocorrelation functions of the mean-subtracted $\varepsilon_{zz}$, $\varepsilon_{yy}$, and $\varepsilon_{xx}$ fields at applied strains of 0, $0.146\%$, and $0.245\%$. The progressively faster decay with applied strain indicates a reduction in the spatial range of the heterogeneous strain field.}
  \label{fig:Figure_3}
\end{figure*}

To directly probe the spatial range of the heterogeneous strain field, we performed strain-dependent HR-EBSD mapping under three macroscopic elastic strain states (details in Supplementary Note 7). A representative Kikuchi pattern used in the cross-correlation analysis is shown in Fig.~\ref{fig:Figure_3}(a), and maps of the loading-direction strain component $\varepsilon_{11}$ at zero and $0.245\%$ applied strain are shown in Figs.~\ref{fig:Figure_3}(b) and (c). Although spatial fluctuations remain visible in both maps, their long-range coherence is strongly reduced under load. We quantified this change using radially averaged autocorrelation functions of the mean-subtracted strain fields. At zero applied strain, the autocorrelations of $\varepsilon_{11}$, $\varepsilon_{22}$, and $\varepsilon_{33}$ remain positive over micrometer-scale distances. With increasing strain, all three functions decrease in amplitude and approach zero over progressively shorter distances [Figs.~\ref{fig:Figure_3}(d)--(f)]. The contraction occurs across multiple tensor components rather than only in the directly loaded direction, indicating a reconfiguration of the heterogeneous elastic field rather than a simple addition of homogeneous strain. The HR-EBSD result provides direct real-space evidence that the range of
dislocation-associated strain disorder decreases with applied strain.

The autocorrelation functions of the strain in GaN are expected to decay approximately
logarithmically below the screening distance and approach zero beyond it,
allowing the screening range to be identified from the crossover in the
correlation function~\cite{kaganer2025dislocation}. The present trends are
therefore consistent with enhanced screening of elastic long-range dislocation strain
fields and independently support the decrease inferred from the full XRD
profile. Because HR-EBSD probes the near-surface region whereas XRD and TDTR
sample micrometer-scale volumes, the agreement among these measurements
indicates that the strain-field reconfiguration is not confined to a single probe scale.

Complementary crystal plasticity fast-Fourier transform and phase-field dislocation-dynamics calculations show that loading the bonded GaN/Ti geometry produces a through-thickness resolved-shear-stress gradient capable of reversibly bowing or straightening pinned threading segments (Supplementary Note 8). These calculations establish the mechanical feasibility of changing the three-dimensional line geometry without glide through the crystal or irreversible dislocation reactions. Because the present simulations do not explicitly predict collective correlations, we use them as a microscopic plausibility argument rather than as a direct calculation of the measured screening length.

\subsection{Strain-dependent Raman signatures of local lattice disorder}

\begin{figure*}[t]
  \centering
  \includegraphics[width=0.8\textwidth]{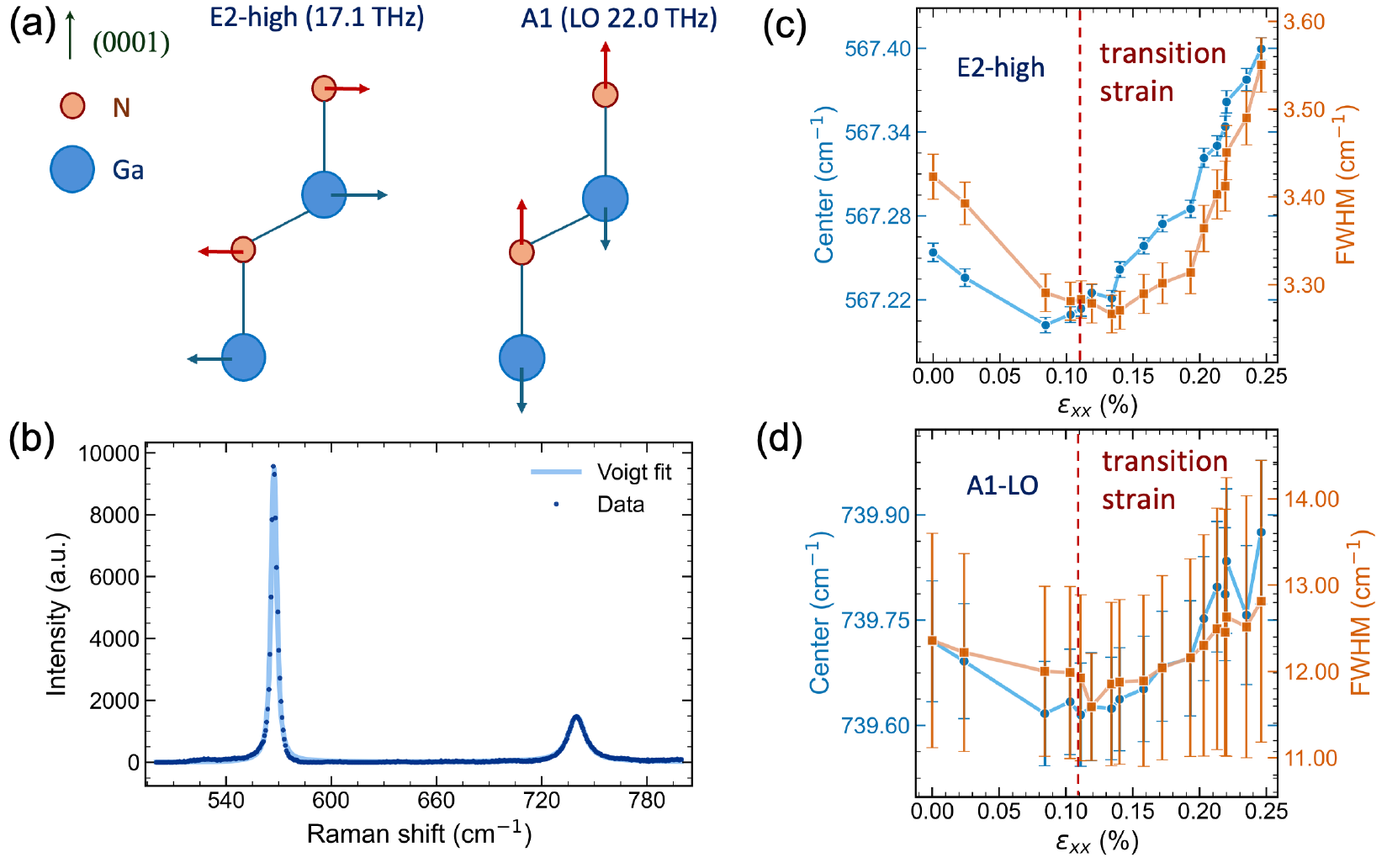}
  \caption{
  \textbf{Strain-dependent Raman spectroscopy of c-plane GaN.}
  (a) The two Raman-active modes probed in GaN.
  (b) The raw Raman spectrum taken at zero strain.
  (c) Evolution of the peak position and linewidth of the $E_{2}^{\mathrm{high}}$ mode as a function of strain.
  (d) Evolution of the peak position and linewidth of the $A_{1}(\mathrm{LO})$ mode as a function of strain.
  }
  \label{fig:Figure_4}
\end{figure*}

To further probe the local lattice response associated with strain-field reconfiguration, we performed \textit{in situ} micro-Raman spectroscopy under the same uniaxial loading (see Supplementary Note 9). Raman scattering is sensitive not only to average elastic strain through phonon-frequency shifts, but also to local disorder and microstrain through changes in phonon linewidth. Figure~\ref{fig:Figure_4}(a) illustrates the dominant Raman-active modes analyzed here: the nonpolar in-plane $E_{2}^{\mathrm{high}}$ mode and the polar $A_{1}(\mathrm{LO})$ mode. A representative spectrum acquired in the unstrained state is shown in Fig.~\ref{fig:Figure_4}(b), with both peaks well resolved and accurately described by Voigt-profile fitting. The strain evolution of the extracted peak positions and linewidths is summarized in Figs.~\ref{fig:Figure_4}(c) and (d). The $E_{2}^{\mathrm{high}}$ mode exhibits the clearest response. Its center frequency shows a non-monotonic variation with applied strain, with a shallow minimum near $\sim0.1\%$ strain followed by an upshift at larger strain. The linewidth initially decreases, reaches a minimum near the onset strain identified in the TDTR measurements, and then broadens progressively at higher strain.

The linewidth evolution indicates a change in the distribution of local strain environments sampled within the approximately $1~\mu$m Raman probe volume. The initial narrowing of the $E_{2}^{\mathrm{high}}$ mode is consistent with reduced inhomogeneous broadening as the long-range heterogeneous strain field becomes screened. Its subsequent broadening at larger strain shows that local strain heterogeneity can increase even while the longer-range correlations measured by HR-EBSD contract. Because phonon linewidths can contain several homogeneous and inhomogeneous contributions, we regard the Raman anomaly as corroborative rather than as a unique structural identification of a particular dislocation arrangement.

The stronger sensitivity of the $E_{2}^{\mathrm{high}}$ mode is physically reasonable because it primarily involves in-plane vibrations of the Ga and N sublattices and therefore couples strongly to in-plane strain. By comparison, the $A_{1}(\mathrm{LO})$ mode, whose atomic motion is largely polarized along the $c$ axis, shows weaker shifts and larger uncertainty [Fig.~\ref{fig:Figure_4}(d)]. The proximity of the $E_{2}^{\mathrm{high}}$ linewidth minimum to the onset of the thermal-conductivity enhancement provides an independent local spectroscopic signature of a change in the strain distribution. Taken together, TDTR, HRXRD, HR-EBSD, and Raman consistently indicate that modest elastic strain shortens the spatial range of dislocation-associated strain disorder, reducing its effective phonon-scattering strength and enhancing heat conduction.
\section{Conclusion}

In summary, we demonstrate that modest elastic strain can reversibly tune thermal transport in defect-rich GaN by modifying dislocation-associated strain fields. A 23\% enhancement of thermal conductivity is achieved under only 0.21\% uniaxial strain. HRXRD reveals pronounced narrowing and evolution of the (0002) reflection, while HR-EBSD directly shows that the spatial autocorrelations of multiple strain components contract under loading. Raman spectroscopy provides a concurrent local signature through the non-monotonic $E_{2}^{\mathrm{high}}$ linewidth. Micromechanical and phase-field calculations further show that the applied stress gradient can reversibly change the geometry of pinned threading segments without requiring irreversible dislocation reactions. Together, these results indicate that applied strain enhances screening of long-range heterogeneous strain fields and reduces their effective phonon-scattering strength. Our work establishes defect-associated strain correlations as a tunable degree of freedom for controlling heat transport in crystalline materials.

\begin{acknowledgments}
We thank James Speck, Youli Li, Vladimir Kaganer, Daniel Gianola, Kirk Fields, and Rachel Schoeppner for helpful discussions. This work is based on research supported by the Office of Naval Research under award number N00014-22-1-2262 and the Army Research Office under award number W911NF2310188. The authors acknowledge the use of the Quantum Structures Facility and the Microscopy \& Microanalysis Facility within the California NanoSystems Institute, supported by the University of California, Santa Barbara (UCSB) and the University of California, Office of the President, as well as the UCSB Materials Research Laboratory Shared Experimental Facilities, which are supported by the NSF MRSEC Program under award number DMR-2308708. 

\end{acknowledgments}
\newpage

\bibliography{references.bib}

\end{document}


\title{Supplementary Information: Reversible Modulation of Thermal Conductivity in GaN through Strain-Field Screening around Dislocations}

\author{Shantal Adajian}
\affiliation{Department of Mechanical Engineering, University of California, Santa Barbara, CA 93106, USA}

\author{Fanghao Zhang}
\affiliation{Department of Mechanical Engineering, University of California, Santa Barbara, CA 93106, USA}

\author{Andrew Christensen}
\affiliation{Department of Mechanical Engineering, University of California, Santa Barbara, CA 93106, USA}

\author{Zeyu Xiang}
\affiliation{Department of Mechanical Engineering, University of California, Santa Barbara, CA 93106, USA}

\author{Tanay Tak}
\affiliation{Materials Department, University of California, Santa Barbara, CA 93106, USA}

\author{Nicholas Fuchs-Lynch}
\affiliation{Materials Department, University of California, Santa Barbara, CA 93106, USA}

\author{Szilvia Kalácska}
\affiliation{Laboratoire Georges Friedel, CNRS, Mines Saint-Étienne, 42023 Saint-Étienne, France}

\author{Nicolò M. della Ventura}
\affiliation{Materials Department, University of California, Santa Barbara, CA 93106, USA}

\author{Miguel Zepeda-Rosales}
\affiliation{Materials Research Laboratory, University of California, Santa Barbara, CA 93106, USA}

\author{Samantha Daly}
\affiliation{Department of Mechanical Engineering, University of California, Santa Barbara, CA 93106, USA}

\author{Irene J. Beyerlein}
\affiliation{Materials Department, University of California, Santa Barbara, CA 93106, USA}

\author{Bolin Liao}
\email{bliao@ucsb.edu}
\affiliation{Department of Mechanical Engineering, University of California, Santa Barbara, CA 93106, USA}

\renewcommand{\thefigure}{S\arabic{figure}}
\renewcommand{\thetable}{S\Roman{table}}
\renewcommand{\theequation}{S\arabic{equation}}                           
\maketitle

\section*{Supplementary Methods}

\subsection{Computational Methods}
Density functional theory (DFT) calculations were accomplished
via the Vienna ab initio simulation package (VASP)~\cite{kresse1996efficiency,kresse1996efficient} based on
the projected augmented wave pseudopotentials~\cite{blochl1994projector}.
Structural optimization was performed using the PBEsol exchange-correlation functional~\cite{perdew2008restoring}.
Plane-wave cutoff energy convergence and \textit{k}-point convergence  were firstly tested to ensure the lattice parameters were well relaxed, yielding optimal values of 550 eV and a $\Gamma$-centered $10\times10\times6$ grid, respectively.
The energy and force convergence criteria were set to $1\times10^{-7}$ eV and $1\times10^{-3}$ eV/\AA, respectively.

For thermal conductivity calculations, the anharmonic third-order IFCs were computed with the $4\times4\times3$ supercells and $1\times1\times1$ \textit{k}-point grids.
The cutoff for neighboring interactions was set as five nearest neighbor, and the same distance was considered for different strains.
The lattice thermal conductivity $k_{ph}$ was then determined by iteratively solving the phonon Boltzmann transport equation (BTE) using the ShengBTE package~\cite{li2014shengbte}, following a convergence test of the \textit{q}-mesh sampling in phonon momentum space.

The phonon-scattering term related to dislocations was approximated via the Carruthers expression~\cite{carruthers/1959/scattering,carruthers1961theory}.

A crystal plasticity fast-Fourier transform (CP-FFT) model was used to determine the heterogeneous stress field that drives the smaller-length-scale phase-field dislocation-dynamics (PFDD) calculations. CP-FFT is a full-field micromechanical method that has been applied to deformation and strain localization in a range of crystalline materials~\cite{lebensohn2012elasto,ahmadikia2021effect}. The model contains GaN bonded to a Ti region three times as thick, with traction-free regions surrounding the exposed GaN surfaces. The basal planes of the two hexagonal materials were joined at the interface, and the Ti was loaded in uniaxial tension to $0.2\%$ strain at a nominal strain rate of $10^{-3}~\mathrm{s^{-1}}$. The calculation produces a through-thickness gradient of resolved shear stress on the GaN pyramidal type-II $\langle c+a\rangle$ slip system, with the largest values near the GaN/Ti interface and decreasing stress toward the free surface.

PFDD was then used to determine how discrete $\langle c+a\rangle$ threading segments can evolve under this stress field. PFDD predicts minimum-energy dislocation pathways under applied stress or temperature and has been used to describe dislocation motion and reactions in both homogeneous and heterogeneous crystals~\cite{beyerlein2016understanding,mianroodi2015atomistically,XU2022105031,albrecht2020phase}. Representative GaN cells were $49~\mathrm{nm}$ thick and either $312$ or $624~\mathrm{nm}$ wide. Threading lines extended between the GaN/Ti interface and the free surface, and the order parameter describing slip was allowed to evolve only within GaN. Straight and initially bowed configurations were considered separately to determine how the response depends on the pre-existing line geometry. The elastic constants and lattice parameters were taken from experimental data, while the generalized stacking-fault energies and Burgers-vector parameters were taken from density-functional-theory and atomistic results for HCP Ti and wurtzite GaN~\cite{albrecht2020phase,chen2015core,simmons1971single}.

\subsection{Experimental Methods}

\subsubsection{Strain-dependent TDTR}

In situ strain-dependent TDTR measurements were performed by adapting a Razorbill Instruments FC100 uniaxial piezoelectric strain cell for optical pump--probe access. The FC100 strain cell provided the mechanical loading, while a TDTR-compatible mounting configuration was designed to expose the aluminum transducer surface during measurement. A ``dog bone''-shaped titanium (Ti) strain platform was used to transfer uniaxial tensile strain to the bonded GaN sample. The platform can be elastically loaded up to 0.35\% tensile strain along the loading direction~\cite{park/2020/rigid}. The strain platform was fabricated from commercially pure Grade 2 titanium foil purchased from Razorbill Instruments (SS40TG), with a thickness of \(100 \pm 5~\mu\mathrm{m}\). The Ti foil was annealed and tension leveled, and the width of the center section of the platform was 0.6 mm.

GaN samples studied in this work were made by hydride vapor phase epitaxy (MTI Corporation) with a thickness of \(350~\mu\mathrm{m}\) and an initial dislocation density smaller than \(5 \times 10^5~\mathrm{cm}^{-2}\). The GaN samples were then mechanically thinned from the backside to a final thickness of \(t = 15 \pm 3\,\mu\mathrm{m}\) using a precision mechanical polisher (Allied High Tech Multiprep System), producing a smooth surface suitable for bonding and ensuring effective strain transfer to the top surface of the sample.

For the TDTR measurement, an \(80~\mathrm{nm}\) aluminum transducer was deposited on the polished surface of each sample using electron-beam evaporation. The GaN samples were adhered to the Ti strain platform using EP62-1LPSP epoxy (Masterbond). The epoxy was applied sparingly to achieve a post-cure thickness of \(1\)--\(3\,\mu\mathrm{m}\). Each sample was placed aluminum side up using MiTeGen polymer-tipped tweezers and gently pressed into place. The platform was then heated to \(135\,^{\circ}\mathrm{C}\) for 4 hours to fully cure the adhesive.

The bonded Ti platform was mounted to the FC100 strain cell (Razorbill Instruments) using Stycast 2850FT epoxy with Catalyst 9 (Emerson \& Cuming). The epoxy was cured at room temperature for 24 hours to ensure rigid fixation for mechanical loading. Uniaxial stress was applied by actuating the piezoelectric stacks of the FC100 strain cell using the RP100 voltage source (Razorbill Instruments). The applied voltage controlled the displacement of the piezoelectric stacks, which in turn stretched the titanium platform and the bonded GaN sample.

The capacitance between the flexure-mounted plates of the FC100 force sensor was monitored using a Keysight E4980A LCR meter. The LCR meter settings and capacitance-to-force conversion procedure followed the recommendations provided by Razorbill Instruments~\cite{razorbill_ap003,razorbill_quickstart}. For each applied strain state, ten capacitance measurements were recorded and averaged to obtain the representative capacitance value, which was then converted to force using the calibration procedure recommended by Razorbill Instruments. At each nonzero voltage (strain) state, the system was allowed to stabilize before TDTR acquisition. The measured capacitance values were reproducible for the same applied voltage across repeated cycles, indicating repeatable loading. 

The maximum accessible strain before fracture varied between samples, typically ranging from \(0.2~\mathrm{\%}\) to \(0.29~\mathrm{\%}\). This variation is expected because the fracture strain depends sensitively on the final sample thickness, local thinning uniformity, and bonding geometry. The GaN thickness was typically measured after the strain-dependent TDTR experiments to avoid introducing additional surface damage or defects before loading. As a result, the loading protocol was chosen conservatively to avoid prematurely breaking the sample before completing the TDTR measurements. Voltage increments were selected to produce capacitance changes larger than the uncertainty of the capacitance measurement, while still limiting the risk of fracture. Multiple measurements were therefore taken within each loading cycle, and selected voltage states were revisited in subsequent cycles to confirm repeatability while preserving the sample.

\subsubsection{Cathodoluminescence}
Cathodoluminescence (CL) measurements were performed using a Gatan MonoCL4 attached to a field emission Thermo Fisher Apreo C scanning electron microscope (SEM). CL measurements were performed in panchromatic mode to improve the signal of photons detected. Threading dislocations of the GaN sample were identified using CL as they appear as dark spots (non-radiative regions) in CL micrographs. Further, the origin of these TDs was determined to be the GaN substrate (rather than damaged surfaces), as their positions and density stayed constant when varying the SEM beam voltage to change the interaction depth of the incident electrons.

\subsubsection{High-resolution x-ray diffraction}
High-resolution X-ray diffraction (HRXRD) measurements were performed using a Rigaku SmartLab diffractometer equipped with a Cu X-ray source, a Ge $(220)\times 2$ incident-beam monochromator, and a HyPix-2000 detector. The incident beam was defined using a slit opening of $1~\mathrm{mm} \times 5~\mathrm{mm}$, with the $1~\mathrm{mm}$ dimension oriented vertically and the $5~\mathrm{mm}$ dimension oriented horizontally. Rocking-curve measurements were collected from the GaN $(0002)$ reflection. For each applied strain state, the sample angle $\omega$ was scanned through the GaN $(0002)$ Bragg condition while recording the diffracted intensity.

\subsubsection{High-resolution electron backscatter diffraction}

High-resolution electron backscatter diffraction (HR-EBSD) was used to map relative elastic-strain fields in GaN under applied macroscopic strains of 0, $0.146\%$, and $0.245\%$. Measurements were performed using a Thermo Fisher Scientific FEI Teneo LoVac field-emission-gun scanning electron microscope (FEG-SEM) equipped with an EDAX Velocity EBSD system. Kikuchi backscatter patterns were acquired \textit{in situ} under load with the sample tilted to $65^\circ$. The dog-bone platform was pin-loaded under displacement control using a Kammrath \& Weiss tension--compression module mounted on the SEM stage. End-to-end displacement and force were measured using an integrated linear variable differential transformer (LVDT) extensometer and a $1~\mathrm{kN}$ load cell, respectively. The strain in the dog-bone gauge section, where the GaN-bearing substrate was bonded, was independently verified from secondary-electron images. Kikuchi patterns were acquired at an accelerating voltage of $20~\mathrm{kV}$, a beam current of $6.4~\mathrm{nA}$, a step size of $50~\mathrm{nm}$, and $2\times2$ detector binning. A separate region of the GaN surface was mapped at each applied strain state.

At each scan point, shifts within regions of interest of the Kikuchi pattern were determined by cross-correlation with a reference pattern and converted to the local elastic-deformation tensor. Spatially uniform offsets were removed to isolate spatial variations in the elastic-strain field. For each normal strain component, the mean-subtracted two-dimensional strain map was autocorrelated, and the resulting two-dimensional autocorrelation was radially averaged to obtain the correlation function as a function of separation $r$. Identical map dimensions and analysis procedures were used for all applied strain states. This analysis follows the HR-EBSD strain-autocorrelation approach used to characterize dislocation screening ranges in GaN~\cite{kaganer2025dislocation}.
\subsubsection{Raman spectroscopy}
Raman spectroscopy was performed using a Confocal Raman and Photoluminescence system based on a Horiba Jobin Yvon T64000 open-frame confocal microscope with a triple monochromator and a liquid-nitrogen-cooled CCD array detector. The instrument is equipped for Raman and photoluminescence spectroscopy and includes 10$\times$, 50$\times$, 50$\times$ ultra-long-working-distance, and 100$\times$ objectives.
For the measurements reported here, Raman spectra were collected using a 514 nm excitation laser, a 50$\times$ VIS ultra-long-working-distance objective, and an 1800 gr/mm grating in single-monochromator mode. Spectra were acquired over the range 400--900 cm$^{-1}$ using an acquisition time of 5 s and 10 accumulations per spectrum. The detector was operated in signal readout mode with binning set to 1, and the CCD temperature was approximately $-130~^\circ$C. The laser filter was set to 100\%. No autofocus, autoexposure, denoising, dark correction, spike filtering, or instrument processing was applied during acquisition.

\subsection{Use of Large Language Model}
The authors acknowledge the use of ChatGPT (OpenAI, GPT-5.6) as an assistive tool in the preparation of this manuscript. The model was used to support drafting and editing of text, improving clarity and organization, and facilitating literature-oriented discussions. All scientific content, data analysis, interpretations, and conclusions were developed, verified, and validated by the authors. The authors take full responsibility for the accuracy and integrity of the work.

\clearpage

\section*{Supplementary Note 1: Finite Element Modeling of Strain Generation and Transmission}

\begin{figure}[ht]
  \centering
  \includegraphics[width=\linewidth]{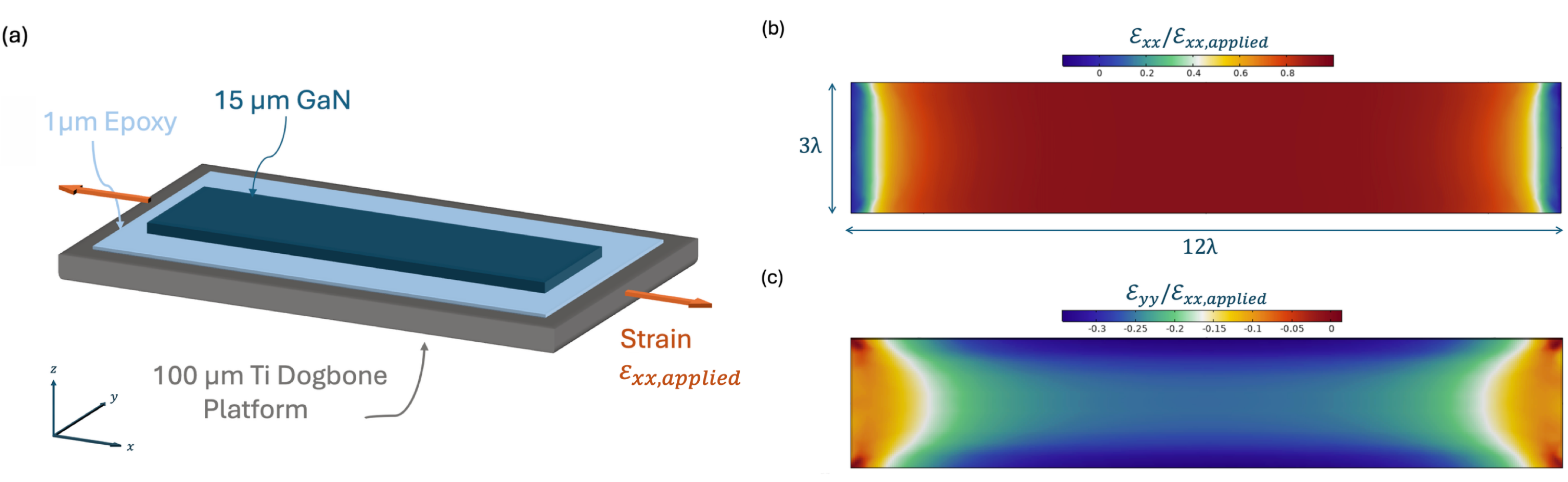}
  \caption{
  \textbf{Finite-element simulation of strain transmission from the platform to a long GaN sample.}
  (a) Simulation geometry used to model strain transfer from the tensile platform to the sample.
  (b) Longitudinal strain $\varepsilon_{xx}$ at the top surface of the sample.
  (c) Transverse strain $\varepsilon_{yy}$ at the top surface of the sample.
  The simulation was based on the method reported in Ref.~\cite{park/2020/rigid}.
  }
  \label{fig:Comsol Data}
\end{figure}

Finite element simulations were performed in \textsc{COMSOL Multiphysics} (Structural Mechanics Module) to quantify strain generation in the titanium platform and strain transmission to a bonded GaN specimen. The platform geometry follows the rigid bowtie-neck design of Park \textit{et al.}~\cite{park/2020/rigid}, in which uniaxial deformation is concentrated within the neck region while the outer tabs are clamped by a uniaxial stress cell. The simulation setup and geometry are shown in Fig.~\ref{fig:Comsol Data}(a). The applied platform strain was defined as
\begin{equation}
\varepsilon_{xx}^{\mathrm{applied}}=\frac{\Delta x}{l_{\mathrm{eff}}},
\end{equation}
where $\Delta x$ is the imposed end displacement and $l_{\mathrm{eff}}$ is the effective neck length.
The GaN specimen was modeled as anisotropic hexagonal (wurtzite) using room-temperature elastic constants from the Ioffe Institute NSM database~\cite{ioffe_GaN_mechanical}, with stiffness tensor
\begin{equation}
\mathbf{C}_{\mathrm{GaN}}=
\begin{pmatrix}
390 & 145 & 106 & 0 & 0 & 0\\
145 & 390 & 106 & 0 & 0 & 0\\
106 & 106 & 398 & 0 & 0 & 0\\
0 & 0 & 0 & 105 & 0 & 0\\
0 & 0 & 0 & 0 & 105 & 0\\
0 & 0 & 0 & 0 & 0 & 122.5
\end{pmatrix}
\ \mathrm{GPa},
\end{equation}
with density $\rho_{\mathrm{GaN}}=6150~\mathrm{kg\,m^{-3}}$. 

The specimen was bonded using EP62-1LPSP epoxy of thickness $d=2\,\mu\mathrm{m}$ and properties $E_{\mathrm{epoxy}}=3.6~\mathrm{GPa}$, $\nu_{\mathrm{epoxy}}=0.30$, and $\rho_{\mathrm{epoxy}}=1530~\mathrm{kg\,m^{-3}}$, giving shear modulus
\begin{equation}
G_{\mathrm{epoxy}}=\frac{E_{\mathrm{epoxy}}}{2(1+\nu_{\mathrm{epoxy}})}\approx 1.38~\mathrm{GPa}.
\end{equation}
All interfaces were assumed rigidly bonded, which is consistent with the experimental fact that no delamination or de-bonding were observed during strain cycling.

Strain transmission from the Ti platform to the GaN sample was analyzed using the strain transfer length formalism of Park \textit{et al.}~\cite{park/2020/rigid}. For c-plane GaN strained along an in-plane direction, the relevant elastic modulus entering the transmission length is
\begin{equation}
C_{\mathrm{eff}} = C_{11} - \frac{C_{13}^{2}}{C_{33}},
\end{equation}
which accounts for elastic coupling between in-plane and out-of-plane deformation. Using the above stiffness constants gives
\begin{equation}
C_{\mathrm{eff}} \approx 362~\mathrm{GPa}.
\end{equation}
The corresponding transmission length is therefore
\begin{equation}
\lambda=\sqrt{\frac{C_{\mathrm{eff}}\,t\,d}{G_{\mathrm{epoxy}}}} \approx 88\,\mu\mathrm{m}.
\end{equation}
Since $L\gg\lambda$, full longitudinal strain transmission is expected in the central region of the specimen. As confirmed in Fig.~\ref{fig:Comsol Data}(b), the ratio $\varepsilon_{xx}^{\mathrm{GaN}}/\varepsilon_{xx}^{\mathrm{Ti}}$ reaches unity away from the specimen ends, with strain relaxation confined to a distance of order $\lambda$ near the edges.
In addition to the longitudinal strain transfer, Fig.~\ref{fig:Comsol Data}(c) shows the transverse strain response $\varepsilon_{yy}^{\mathrm{GaN}}/\varepsilon_{xx}^{\mathrm{Ti}}$. As discussed by Park \textit{et al.}~\cite{park/2020/rigid}, transverse strain is governed by the sample Poisson ratio near the specimen edges, while toward the center it becomes increasingly controlled by the platform Poisson ratio. Because the present specimen width ($w=250\,\mu\mathrm{m}$) is comparable to the transmission length ($\lambda \approx 88\,\mu\mathrm{m}$), the transverse strain exhibits nonuniformity across the bonded region, consistent with the intermediate-width regime described in Ref.~\cite{park/2020/rigid}.

\clearpage

\section*{Supplementary Note 2: TDTR Data Analysis and Other Measurements}

The TDTR analysis follows a multilayer heat diffusion model similar to that used in previous work from our group~\cite{plunkett2023blending,niyikiza2025thermal}, where the measured thermoreflectance response is fitted using a solution to the Fourier heat conduction equation for a layered structure~\cite{jiang2018tutorial}. In TDTR, the thermal conductivity is obtained by fitting the time dependence of the ratio of the in-phase and out-of-phase lock-in signals, $-V_{\mathrm{in}}/V_{\mathrm{out}}$, or equivalently the TDTR phase, using a multilayer thermal diffusion model.
For the present GaN measurements, the sample was modeled as an Al transducer layer on GaN, with heat flow across the Al/GaN interface and into the GaN layer. The Al layer serves as the optical transducer, absorbing the modulated pump beam and converting the transient temperature response into a change in reflectance measured by the probe beam. The unknown or fitted parameters include the GaN thermal conductivity and the Al/GaN interfacial thermal conductance, while literature or independently measured values were used for the remaining parameters, including the Al thickness, Al thermal properties, optical spot sizes, and volumetric heat capacities. The pump and probe beams were focused using a 10$\times$ objective lens. The measured pump and probe diameters were approximately $35~\mu\mathrm{m}$ and $9~\mu\mathrm{m}$. The pump beam was modulated at $10~\mathrm{MHz}$.

The TDTR thermal model solves the heat diffusion equation in cylindrical coordinates,

\begin{equation}
C\frac{\partial T}{\partial t}
=
\eta k_z
\frac{1}{r}
\frac{\partial}{\partial r}
\left(
r\frac{\partial T}{\partial r}
\right)
+
k_z
\frac{\partial^2 T}{\partial z^2},
\end{equation}
where $C$ is the volumetric heat capacity, $T$ is temperature, $k_z$ is the cross-plane thermal conductivity, and $\eta = k_r/k_z$ accounts for thermal anisotropy. The multilayer temperature response is calculated in the frequency domain using a transfer-matrix approach, where interfacial thermal conductance is included through a thermal boundary resistance between adjacent layers.

To evaluate which parameters most strongly affect the extracted GaN thermal conductivity, a sensitivity analysis was performed. The sensitivity of the TDTR phase $\Phi$ to a parameter $x$ was defined as the logarithmic derivative,

\begin{equation}
S_x
=
\frac{\partial \ln \Phi}{\partial \ln x}.
\end{equation}

\begin{figure}[ht]
  \centering
  \includegraphics[width=\linewidth]{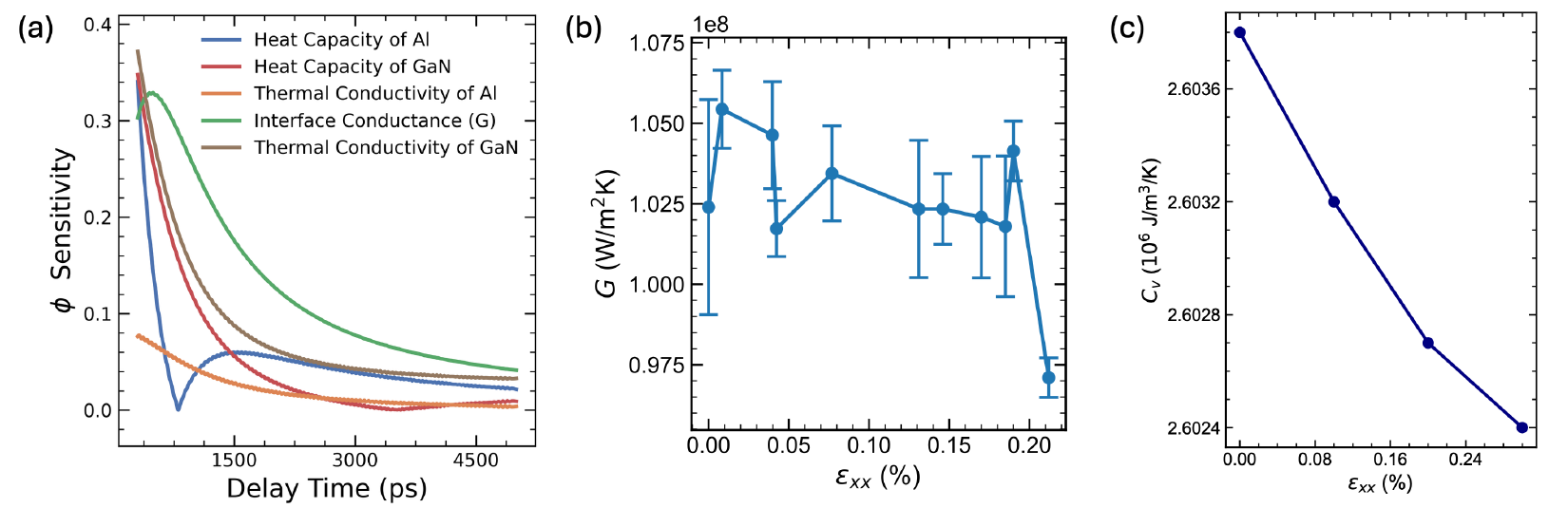}
  \caption{
  \textbf{TDTR sensitivity analysis.}
  Sensitivity of the TDTR phase to the main thermal parameters used in the multilayer heat diffusion model. (a) The phase response is most sensitive to the GaN thermal conductivity in the selected fitting range, while the sensitivities to the Al transducer properties and Al/GaN interfacial thermal conductance are smaller. (b) Fitted Al/GaN interfacial thermal conductance as a function of applied tensile strain. (c) Calculated strain dependence of GaN volumetric heat capacity using DFT.
  }
  \label{fig:tdtr_sensitivity}
\end{figure}

This analysis follows the common TDTR protocol, where the phase sensitivity is evaluated with respect to fitting parameters such as interfacial conductance and film thermal conductivity. The resulting sensitivity curves are shown in Fig.~\ref{fig:tdtr_sensitivity}(a). The TDTR signal is most sensitive to the GaN thermal conductivity over the selected fitting window, while the sensitivity to the Al/GaN interfacial thermal conductance and Al thermal properties is comparatively smaller. Therefore, the strain-dependent trend extracted from the TDTR data is primarily governed by changes in the GaN thermal conductivity rather than artifacts from the transducer layer or interface.

Because mechanical strain is applied to the full sample stack, possible strain-induced changes in the Al transducer were also considered. The Al layer is thin compared with the GaN substrate/layer and acts primarily as an optical transducer. Therefore, even if the Al thermal conductivity or heat capacity changes slightly under the small applied strains used here, its impact on the fitted GaN thermal conductivity is limited. This is evidenced by the lower sensitivity to Al heat capacity and thermal conductivity shown in Fig.~\ref{fig:tdtr_sensitivity}(a).

We also examined whether the Al/GaN interfacial thermal conductance changes systematically with applied strain. The fitted Al/GaN interfacial conductance shows no strong monotonic dependence on strain, indicating that the measured strain dependence is not dominated by changes in thermal boundary conductance. The fitted interface values remain comparatively stable with strain as shown in Fig.~\ref{fig:tdtr_sensitivity}(b).

Finally, the possible effect of strain-dependent GaN volumetric heat capacity was evaluated.  First-principles density functional theory and density functional perturbation theory calculations predict only a small change in the GaN volumetric heat capacity over the applied strain range, with a maximum relative change of approximately $-0.054\%$. This change is much smaller than the measured strain-dependent change in thermal conductivity and is therefore insufficient to explain the TDTR trend. The calculated strain dependence of the GaN heat capacity is shown in Fig.~\ref{fig:tdtr_sensitivity}(c).

Overall, the sensitivity analysis, fitted Al/GaN interfacial conductance, and calculated strain-dependent GaN heat capacity support the conclusion that the measured TDTR response reflects a genuine strain-dependent change in the GaN thermal conductivity.

\clearpage

\section*{Supplementary Note 3: Reversibility of the Measured Thermal Conductivity Change}

\begin{figure}[!htbp]
  \centering
  \includegraphics[width=0.5\linewidth]{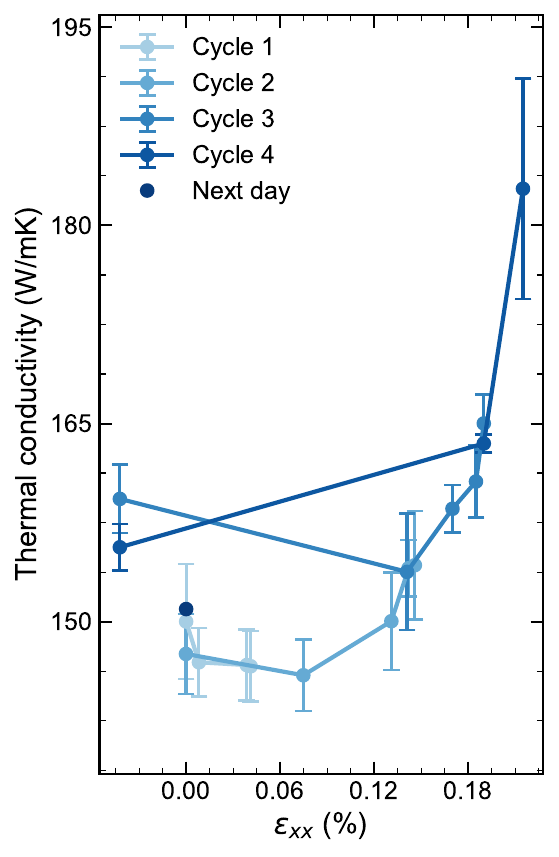}
  \caption{\textbf{Cycling repeatability of strain-dependent TDTR measurements.} Thermal conductivity measured as a function of applied voltage during repeated loading cycles of the piezoelectric strain cell. }
  \label{fig:cycling_data}
\end{figure}
To evaluate the repeatability of the strain-dependent TDTR measurements and the effect of loading history, the sample was cycled through multiple applied-voltage states by actuating the piezoelectric stacks of the FC100 strain cell, as shown in Fig.~\ref{fig:cycling_data}. The plotted dataset combines four repeated loading cycles rather than one continuous voltage sweep. The voltage-to-strain response was nonlinear, with larger capacitance changes at low voltage than at high voltage. Therefore, voltage was used only as the actuation parameter, while strain was determined from the measured capacitance response. We note that after returning the piezo voltage to zero at the beginning of each cycle, the strain-cell capacitance exhibited a small transient offset, typically corresponding to a nominal compressive strain (shown here in Fig.~\ref{fig:cycling_data}), which relaxed back to the zero-strain value over several hours. This behavior is consistent with piezoelectric hysteresis as noted in the manual of the strain-cell rather than irreversible deformation of the GaN sample. This is supported by the fact that the thermal conductivity of the zero-strain state measured the next day after the full relaxation of the piezo stack returned to the original value before strain cycling (Fig.~\ref{fig:cycling_data}).

\clearpage

\section*{Supplementary Note 4: First-principles Simulation of Strain-dependent Thermal Conductivity in GaN}

\begin{figure}[!htbp]
  \centering
  \includegraphics[width=0.9\linewidth]{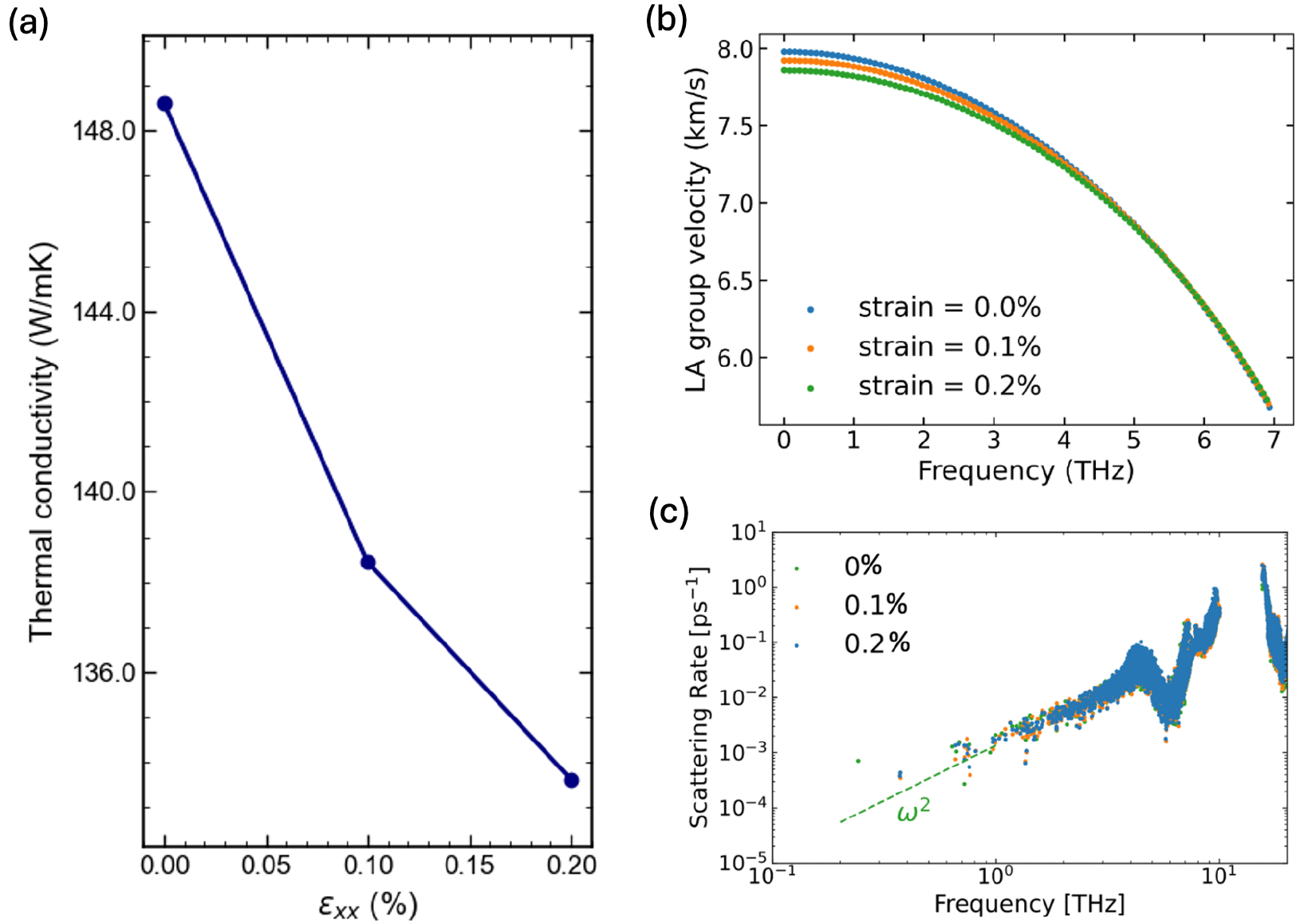}
  \caption{\textbf{First-principles simulation of phonon transport in GaN under different tensile strains.} (a) The calculated cross-plane thermal conductivity in GaN as a function of strain. (b) The group velocity along the c axis of zone-center LA phonons in GaN under different strains. (c) The phonon scattering rates in GaN under different strains.}
  \label{fig:phonon_sim}
\end{figure}

We calculated the cross-plane thermal conductivity of c-plane GaN with a dislocation density of $10^9$ cm$^{-2}$ as a function of tensile strain. The calculated thermal conductivity values as shown in Fig.~\ref{fig:phonon_sim}(a)  corresponding to a roughly 10\% reduction, in sharp contrast to the 23\% increase observed experimentally. As shown in Fig.~\ref{fig:phonon_sim} (b) and (c), the main difference in the computed thermal conductivity stems from the reduced phonon group velocity along the c axis with a tensile strain. This is classical behavior expected in elastically strained covalent crystals.

\clearpage

\section*{Supplementary Note 5: Strain-dependent Thermal Conductivity in GaN with Other Orientations}
\begin{figure}[!htbp]
  \centering
  \includegraphics[width=0.85\linewidth]{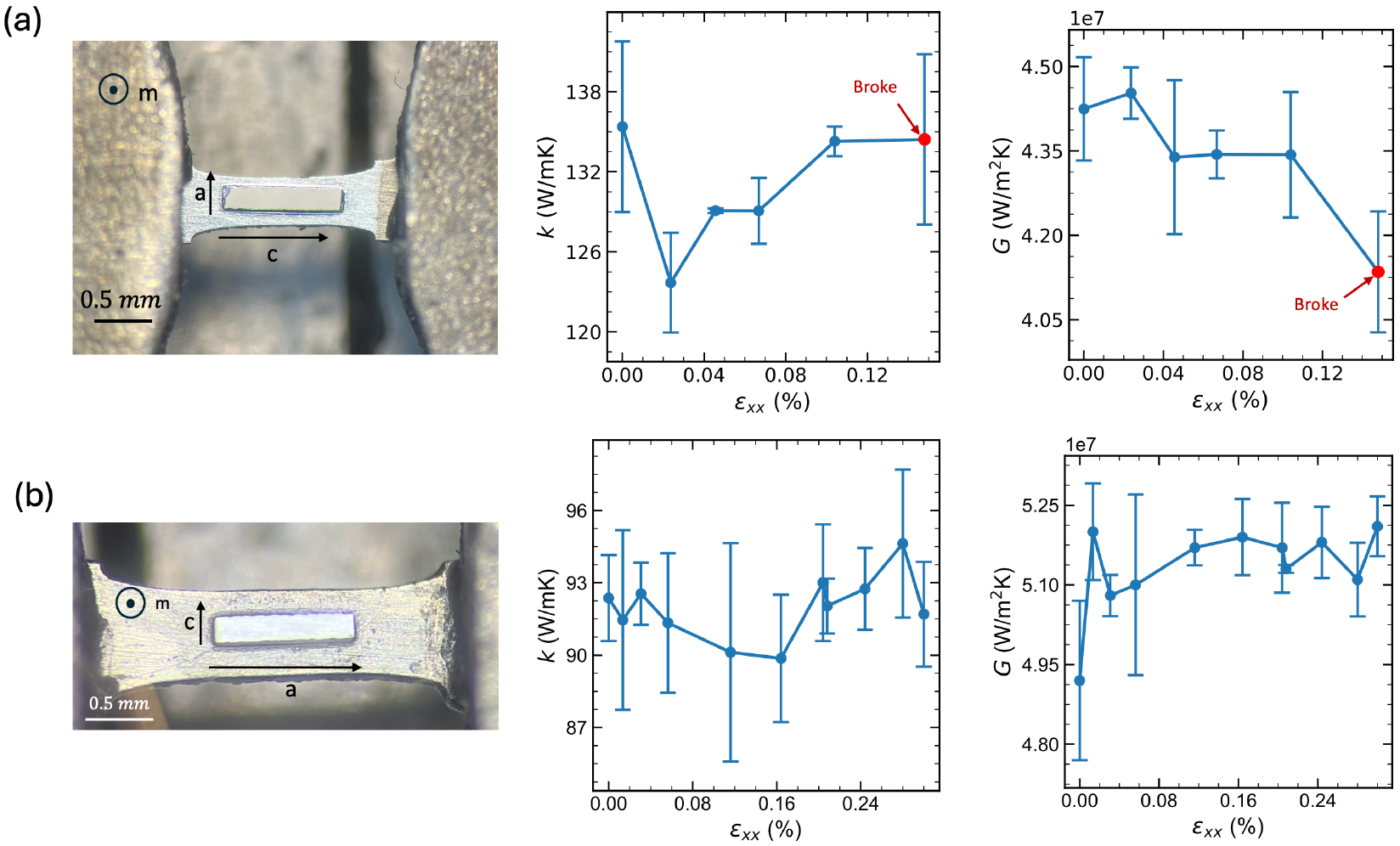}
  \caption{
  \textbf{Strain-dependent thermal conductivity and Al/GaN interfacial thermal conductance of $m$-plane $(10\bar{1}0)$ GaN.}
  Thermal conductivity $\kappa$ and fitted Al/GaN interfacial thermal conductance $G$ are plotted as a function of applied tensile strain for $m$-plane GaN strained along (a) the $c$ direction and (b) the $a$ direction.
  }
  \label{fig:GaN_m_k_G}
\end{figure}

To compare the strain response of different GaN orientations, we also performed TDTR measurements on $m$-plane $(10\bar{1}0)$ GaN under tensile strain along $c$ and $a$ axis, as shown in Fig.~\ref{fig:GaN_m_k_G}. While the main manuscript focuses on strain-dependent thermal conductivity in $c$-plane GaN, the $m$-plane measurements provide an additional comparison of the robustness of the strain response in a nonpolar orientation.

For $m$-plane GaN strained along the $c$ direction [Fig.~\ref{fig:GaN_m_k_G}(a)], the sample fractured near $\sim0.13\%$ applied tensile strain, preventing measurements at larger strains. Despite this limited strain range, the extracted thermal conductivity did not show a drastic change with applied strain for $m$-plane GaN strained along the $c$ direction. The fitted Al/GaN interfacial thermal conductance also remained relatively unchanged with strain, indicating that the measured TDTR response was not dominated by strain-induced changes in the interface.

We also measured $m$-plane GaN strained along the $a$ direction [Fig.~\ref{fig:GaN_m_k_G}(b)]. In this configuration, the strain was applied monotonically in a single loading sequence rather than through repeated cycling because the thermal conductivity showed no clear systematic change with strain. Similar to the $c$-direction loading case, neither the extracted thermal conductivity nor the fitted Al/GaN interfacial thermal conductance exhibited a strong strain dependence.
\clearpage

\clearpage
\section*{Supplementary Note 6: Analysis of HRXRD Measurement}

The measured GaN $(0002)$ rocking curves were analyzed following the dislocation peak-profile framework of Kaganer \textit{et al.}~\cite{kaganer/2005/x-ray}. For each strain state, the rocking curve was centered at the fitted peak maximum, and the intensity was analyzed as a function of the angular deviation from the peak center,

\begin{equation}
\Delta \omega = \omega - \omega_0,
\end{equation}
where $\omega_0$ is the peak position. The resulting profiles were used to evaluate the strain-dependent evolution of the rocking-curve center, full width at half maximum (FWHM), integrated intensity, and peak-tail behavior.

Following Kaganer \textit{et al.}~\cite{kaganer/2005/x-ray}, the diffraction peak profile from threading dislocations can be described using a restricted random dislocation distribution. In this framework, the central part of the peak can be approximately Gaussian, while the peak tails contain information about the strain fields associated with dislocations. For randomly distributed dislocations, the far tails of the rocking curve are expected to approach a $q^{-3}$ asymptotic decay. The full profile can be fitted using the expression:

\begin{equation}
I(\omega)
=
\frac{I_i}{\pi}
\int_0^{\infty}
\exp
\left[
-Ax^2
\ln
\left(
\frac{B+x}{x}
\right)
\right]
\cos(\omega x)
\,dx
+
I_{\mathrm{backgr}},
\end{equation}
where $I_i$ is the integrated intensity of the diffraction peak, $I_{\mathrm{backgr}}$ is the background intensity, and $A$ and $B$ are fitting parameters related to the dislocation density and dislocation correlation range, respectively. The fitting parameters are related to the dislocation structure by

\begin{equation}
A = f \rho b^2,
\end{equation}
and

\begin{equation}
B = g\frac{L}{b},
\end{equation}
where $\rho$ is the threading dislocation density, $b$ is the Burgers vector magnitude, $L$ is the dislocation correlation length, and $f$ and $g$ are dimensionless geometrical factors determined by the diffraction geometry.

Because the GaN $(0002)$ reflection is a symmetric reflection with the scattering vector along the $c$ direction, the rocking-curve broadening is primarily sensitive to screw-type threading dislocations. For this geometry, the relevant Burgers vector is $b_{\mathrm{screw}}=c=0.518$\,nm, and the geometrical factors for screw dislocations are
\begin{equation}
f = \frac{1}{8\pi},
\end{equation}
and
\begin{equation}
g = 2\pi.
\end{equation}
Therefore, the screw-threading-dislocation density was calculated from the fitted parameter $A$ as
\begin{equation}
\rho_{\mathrm{screw}}
=
\frac{A}{f b_{\mathrm{screw}}^2}
=
\frac{8\pi A}{b_{\mathrm{screw}}^2}.
\end{equation}
The dislocation correlation length was calculated from the fitted parameter $B$ as
\begin{equation}
L
=
\frac{B b_{\mathrm{screw}}}{g}
=
\frac{B b_{\mathrm{screw}}}{2\pi}.
\end{equation}
The dimensionless screw-dislocation correlation parameter was then defined as
\begin{equation}
M_{\mathrm{screw}}
=
L\sqrt{\rho_{\mathrm{screw}}}.
\end{equation}
Here, $M_{\mathrm{screw}}$ compares the dislocation correlation length $L$ with the mean screw-dislocation spacing $\rho_{\mathrm{screw}}^{-1/2}$. Larger values of $M_{\mathrm{screw}}$ indicate that the dislocation strain fields remain correlated over a longer distance relative to the mean spacing between screw-type threading dislocations.

\begin{figure}[!htbp]

  \centering

  \includegraphics[width=\linewidth]{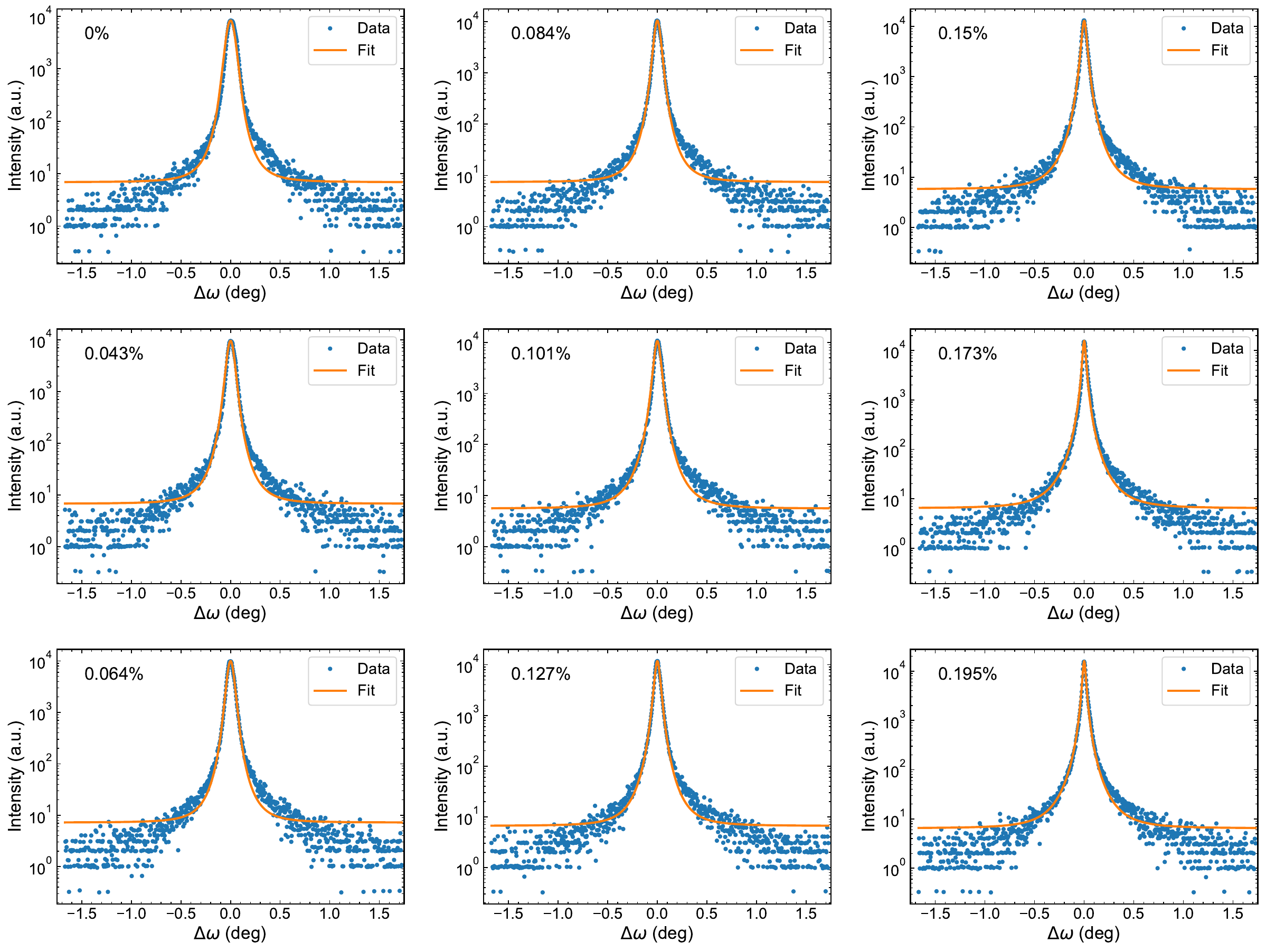}

  \caption{\textbf{Dislocation peak-profile fitting of GaN $(0002)$ rocking curves.} Representative GaN $(0002)$ rocking curves measured at different strain values. The model captures both the central peak region and the extended peak tails, enabling extraction of the fitted parameters $A$ and $B$, which are related to the screw-threading-dislocation density and correlation length, respectively.}

  \label{fig:Eq10_Fit}

\end{figure}

\begin{figure}[!htbp]
  \centering
  \includegraphics[width=0.45\linewidth]{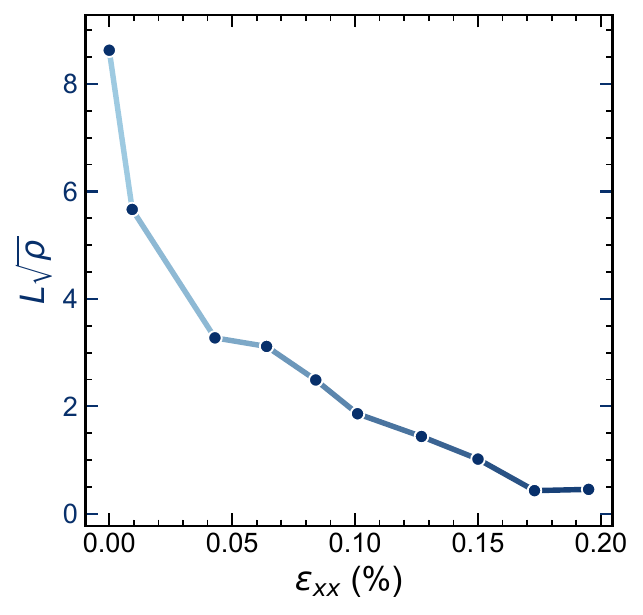}
  \caption{
  \textbf{Strain dependence of the screw-dislocation correlation parameter.}
  The dimensionless parameter $M_{\mathrm{screw}}=L\sqrt{\rho_{\mathrm{screw}}}$ was calculated from the fitted dislocation correlation length $L$ and screw-threading-dislocation density $\rho_{\mathrm{screw}}$ extracted from the GaN $(0002)$ rocking-curve fits. This parameter compares the dislocation correlation length to the mean screw-dislocation spacing and provides an effective measure of the relative range of screw-dislocation strain-field correlations as a function of applied tensile strain.
  }
  \label{fig:M_screw}
\end{figure}

Representative fits of the GaN $(0002)$ rocking curves using the full peak-profile expression are shown in Fig.~\ref{fig:Eq10_Fit}. The model captures the central peak shape and the extended peak tails, allowing the fitted parameters $A$ and $B$ to be extracted at each applied strain state.
While the full peak-profile model has a $q^{-3}$ far-tail asymptote, the measured GaN $(0002)$ profiles under strain do not follow a perfect $q^{-3}$ dependence over the full experimentally accessible tail region. The measured slope evolves from approximately $-3$ toward a shallower value near $-2$ with applied strain, as shown in Fig.~2(e) of the main text. The Kaganer model predicts that the far-tail $q^{-3}$ asymptote is insensitive to the dislocation correlation range; moreover, $q^{-2}$ behavior in a symmetric reflection can also result from finite-thickness or coherent-domain-size broadening~\cite{kaganer/2005/x-ray}. The change in exponent is therefore not interpreted as an independent or unique signature of dislocation screening. Instead, it shows that the relative contributions to the rocking-curve broadening evolve under load. Accordingly, the values of $L$ and $M_{\mathrm{screw}}=L\sqrt{\rho_{\mathrm{screw}}}$ extracted from the full-profile expression are treated as effective comparative metrics rather than absolute dislocation-correlation lengths in a purely ideal asymptotic regime. The strain dependence of $L$ is shown in Fig.~2(e) of the main text, and the strain dependence of $M_{\mathrm{screw}}$ is shown in Fig.~\ref{fig:M_screw}. The direct HR-EBSD autocorrelations in Fig.~3 provide the primary real-space evidence for a contraction of the heterogeneous strain-field range.

\newpage
\clearpage

\section*{Supplementary Note 7: High-Resolution Electron Backscatter Diffraction Analysis}

The EBSD patterns were analyzed using CrossCourt version 4.3 (BLG Vantage, UK). At each measurement point, shifts in multiple regions of interest within the Kikuchi pattern were determined by cross-correlation with a reference pattern. These shifts were used to calculate the local elastic-deformation tensor relative to the selected reference point.

Because the maps were acquired from different regions of the GaN surface, a separate reference pattern was selected independently for each map rather than transferring one reference pattern between maps. Where available, the reference was chosen from a relatively stable, low-kernel-average-misorientation (low-KAM) region near the center of the field of view. This consistent selection criterion was applied to all maps to reduce the influence of local lattice-rotation gradients, pattern distortion near the map boundaries, and isolated high-strain features. Consequently, the reported HR-EBSD strains represent relative intramap variations and should not be interpreted as absolute strain values referenced to a common material point across the different fields of view.

The anisotropic elastic response of wurtzite GaN was treated using the same room-temperature stiffness constants employed in the COMSOL strain-transfer calculations from the Ioffe Institute NSM database~\cite{ioffe_GaN_mechanical},

Before calculating the autocorrelation, the spatial mean of each strain-component
map was subtracted:
\begin{equation}
\delta\varepsilon_{ij}(\mathbf{x})
=
\varepsilon_{ij}(\mathbf{x})
-
\left\langle\varepsilon_{ij}\right\rangle.
\end{equation}
Because the spatial mean is subtracted independently from each EBSD strain map at each applied strain state, the autocorrelation analysis is insensitive to spatially uniform strain offsets associated with the choice of reference pattern. Consequently, the reference pattern need not be selected from the unloaded ($0\%$ strain) condition.

The two-dimensional strain autocovariance was evaluated using
fast-Fourier transform (FFT)-based convolution. For a displacement
$\boldsymbol{\Delta}$, the autocovariance was calculated as

\begin{equation}
C_{ij}(\boldsymbol{\Delta})
=
\frac{
\sum_{\mathbf{x}}
m(\mathbf{x})m(\mathbf{x}+\boldsymbol{\Delta})
\delta\varepsilon_{ij}(\mathbf{x})
\delta\varepsilon_{ij}(\mathbf{x}+\boldsymbol{\Delta})
}{
\sum_{\mathbf{x}}
m(\mathbf{x})m(\mathbf{x}+\boldsymbol{\Delta})
},
\end{equation}
where $m(\mathbf{x})$ is a binary mask equal to unity at valid measurement
points and zero otherwise. The numerator was computed by FFT convolution of
the mean-subtracted strain map with its spatially reversed copy. An equivalent
FFT convolution of the binary mask was used to determine the number of
overlapping valid pixel pairs at each displacement. Displacements containing
fewer than 10 valid pixel pairs were excluded.

The resulting two-dimensional autocovariance was grouped into radial bins
according to $r=|\boldsymbol{\Delta}|$ and averaged over direction, with each
displacement weighted by its number of valid pixel pairs. The radial-bin width
was one EBSD step, corresponding to $50~\mathrm{nm}$. This produced the
orientation-averaged strain autocovariance $C_{ij}(r)$ ~\cite{kaganer2025dislocation}. The same preprocessing and autocovariance procedure was applied to
all strain states. This analysis probes the spatial range of the heterogeneous
elastic-strain field rather than the spatially uniform applied strain and
follows the HR-EBSD strain-correlation approach used to evaluate dislocation
screening in GaN~\cite{kaganer2025dislocation}.
\clearpage

\section*{Supplementary Note 8: Micromechanical and Phase-Field Dislocation-Dynamics Calculations}

\begin{figure}[!htbp]
  \centering
  \includegraphics[
    width=0.65\textwidth,
    keepaspectratio
  ]{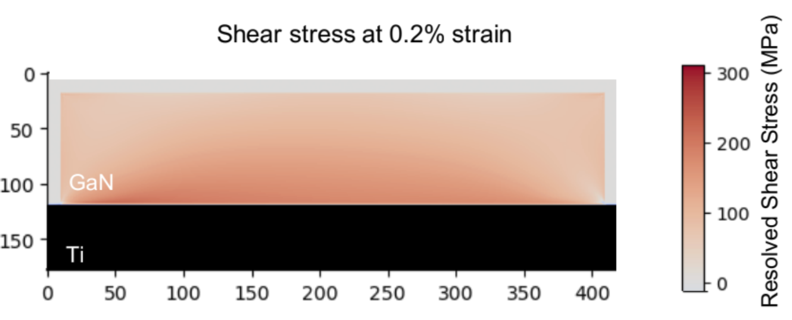}
  \caption{\textbf{Resolved shear stress in the bonded GaN/Ti geometry.} CP-FFT map of the resolved shear stress on the GaN pyramidal type-II $\langle c+a\rangle$ slip system when $0.2\%$ tensile strain is applied to the Ti region. The stress is largest near the GaN/Ti interface and decreases toward the GaN free surface.}
  \label{fig:cpfft_stress}
\end{figure}

The CP-FFT calculation predicts that applying $0.2\%$ tensile strain to the Ti platform generates a resolved-shear-stress gradient through the GaN thickness on the pyramidal type-II $\langle c+a\rangle$ slip system. As shown in Fig.~\ref{fig:cpfft_stress}, the resolved shear stress is largest near the bonded interface and decreases toward the traction-free GaN surface. This inhomogeneous stress field provides a driving force for changes in the shape of threading segments even when their endpoints remain constrained.

\begin{figure}[!htbp]
  \centering
  \makebox[\textwidth][c]{%
    \includegraphics[
      width=1\textwidth,
      keepaspectratio
    ]{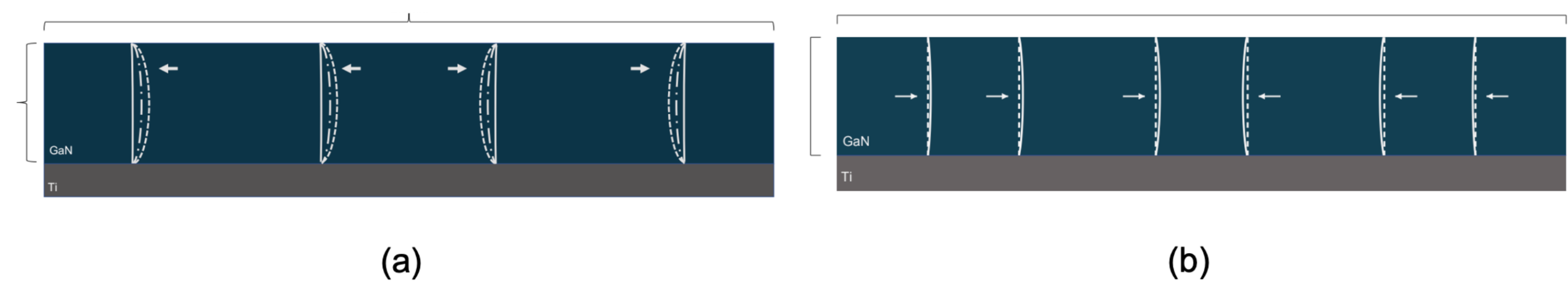}%
  }
  \caption{\textbf{Reversible changes in threading-dislocation line geometry predicted by PFDD.} (a) Bowing of initially straight $\langle c+a\rangle$ threading segments. (b) Straightening of initially bowed segments. Dashed, dash-dotted, and solid curves indicate initial, intermediate, and final configurations as applicable. The response depends on the initial line geometry and local driving direction.}
  \label{fig:pfdd_configurations}
\end{figure}

The PFDD calculations show two limiting responses determined by the initial dislocation geometry. Initially straight threading segments can bow under the calculated stress gradient [Fig.~\ref{fig:pfdd_configurations}(a)], increasing their total line length by at most $3.09\%$ among the configurations tested. Conversely, initially bowed segments can straighten under the opposite local driving direction [Fig.~\ref{fig:pfdd_configurations}(b)], reducing their line length. In both cases the lines remain pinned at the interface and free-surface endpoints, and the applied stresses are insufficient to drive glide through the GaN or irreversible reactions such as annihilation, junction formation, or entanglement. The number of surface-reaching threading dislocations is therefore preserved, consistent with the CL observations and the reversible TDTR response.

These calculations do not explicitly simulate the collective positional correlations of an interacting dislocation ensemble and therefore do not independently predict the HR-EBSD screening distance or the sign of the thermal-conductivity change. Rather, they demonstrate a mechanically plausible, reversible route by which substrate loading can change the three-dimensional geometry and elastic fields of pinned threading dislocations without changing their number or surface positions. A direct comparison with the HR-EBSD data would require calculating strain maps and spatial autocorrelation functions from statistically representative simulated ensembles.

\clearpage

\section*{Supplementary Note 9: Analysis of Raman Spectroscopy}

The Raman spectra were analyzed by fitting the baseline-corrected spectra with a sum of Voigt functions. First, a linear background was estimated from the lowest-intensity points in the selected fitting window and subtracted from the raw spectrum,

\begin{equation}
I_{\mathrm{corr}}(\omega)
=
I_{\mathrm{raw}}(\omega)
-
\left(m\omega + b\right),
\end{equation}
where $I_{\mathrm{raw}}$ is the measured Raman intensity, $\omega$ is the Raman shift, and $m\omega+b$ is the estimated linear background. Negative values after background subtraction were clipped to zero.
The corrected spectrum was then fitted using a sum of Voigt profiles with an additional residual linear baseline,
\begin{equation}
I_{\mathrm{fit}}(\omega)
=
m_r \omega + b_r
+
\sum_{i=1}^{N}
V_i(\omega),
\end{equation}
where $m_r \omega + b_r$ accounts for any remaining linear background in the fitting window and $N$ is the number of Raman peaks included in the fit. Each Raman peak was modeled using a Voigt profile,

\begin{equation}
V_i(\omega)
=
A_i
\frac{
\mathrm{Re}
\left[
w\left(
\frac{(\omega-\omega_i)+i\gamma_i}{\sigma_i\sqrt{2}}
\right)
\right]
}{
\sigma_i\sqrt{2\pi}
},
\end{equation}
where $A_i$ is the peak amplitude scaling factor, $\omega_i$ is the fitted peak center, $\sigma_i$ is the Gaussian broadening parameter, $\gamma_i$ is the Lorentzian broadening parameter, and $w(z)$ is the Faddeeva function. The Gaussian contribution represents inhomogeneous or instrumental broadening, while the Lorentzian contribution captures homogeneous linewidth broadening.
The fitted peak centers $\omega_i$ were used to track Raman peak shifts as a function of applied strain. The fitted linewidth parameters were used to evaluate changes in Raman peak broadening with strain.

\begin{figure}[ht]
  \centering
  \includegraphics[width=\linewidth]{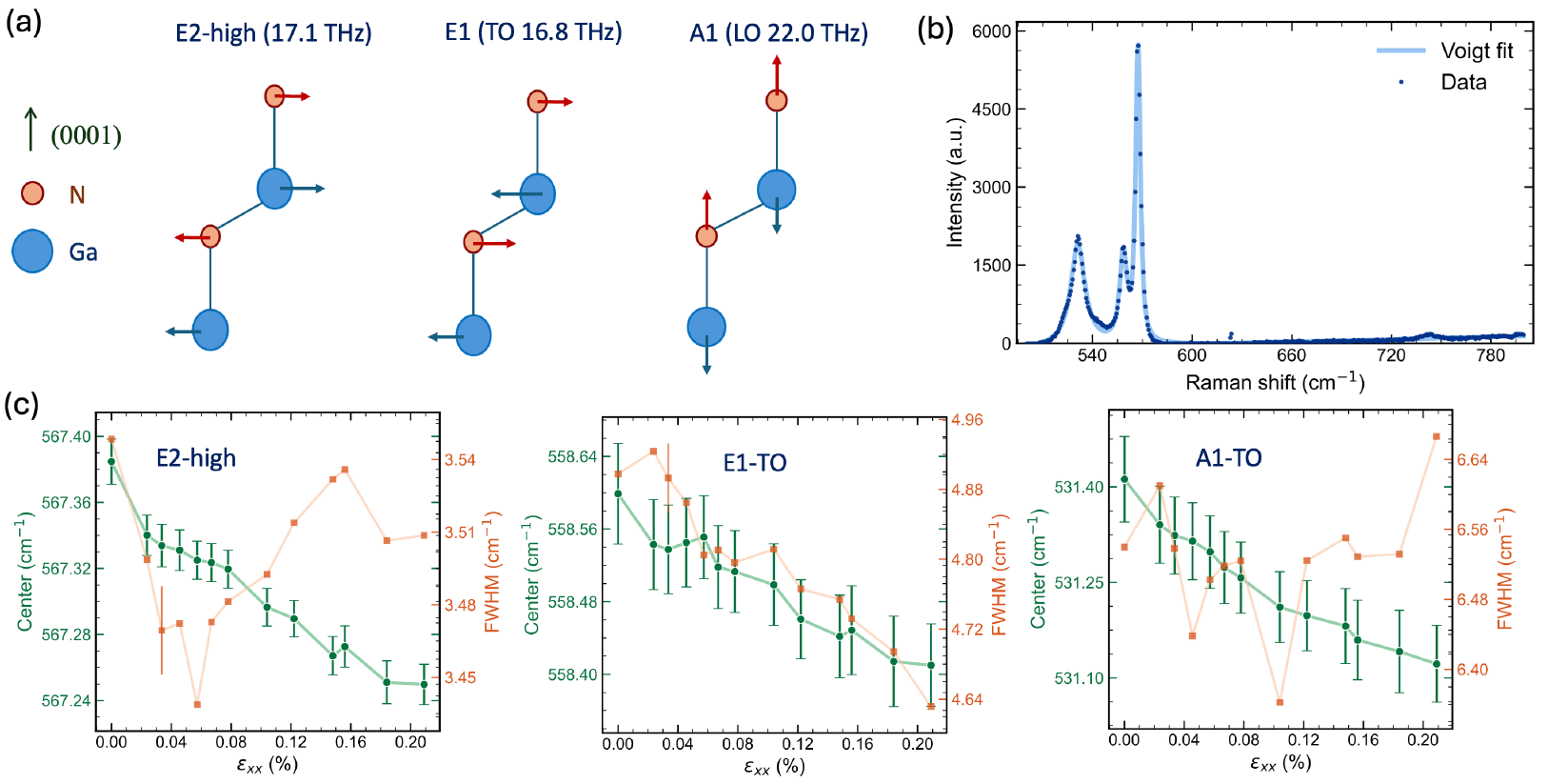}
  \caption{
  \textbf{Strain-dependent Raman spectroscopy of m-plane $(10\bar{1}0)$ GaN under uniaxial tensile strain applied along the $a$ direction. }
  (a) Schematic of the Raman-active phonon modes observed. The atomic displacement directions illustrate the characteristic vibrational motion of each mode.
  (b) Representative baseline-corrected Raman spectrum with Voigt peak fitting. The measured data are shown together with the fitted Voigt line shape used to extract the Raman peak positions and linewidths.
  (c) Strain dependence of the fitted Raman peak centers and FWHM.
  }
  \label{fig:raman_strain}
\end{figure}

Raman measurements were also performed on m-plane $(10\bar{1}0)$ GaN under uniaxial tensile strain applied along the $a$ direction as shown in Fig.~\ref{fig:raman_strain}. Strain applied along the $c$ direction was not pursued for the Raman strain series because the sample geometry was more delicate in this configuration and typically fractured near $\sim 0.13\%$ as shown in Fig.~\ref{fig:GaN_m_k_G}(a). 
\clearpage

\section*{Supplementary References}
\bibliography{references.bib}